\documentclass[12pt, letterpaper]{article}
\usepackage{amsmath}
\usepackage{amsfonts}
\usepackage{amssymb}
\usepackage{graphicx}

\usepackage[colorlinks]{hyperref}
\usepackage[usenames,dvipsnames]{color}
\hypersetup{
     breaklinks=true,
    pdfstartview={FitH},    
    colorlinks=true,       
    linkcolor=magenta,          
    citecolor=Green,        
    filecolor=magenta,      
    urlcolor=Cyan,           
    anchorcolor=green,      
    linktocpage=true
}
\usepackage{color}
\usepackage{amsmath, amssymb, graphics, epsfig, graphicx}
\usepackage{epsf}
\usepackage{epstopdf}
\usepackage {amssymb}
\newcommand{\nc}{\newcommand}
\nc{\ba}{\begin{eqnarray}}
\nc{\ea}{\end{eqnarray}}
\newcommand\be{\begin{equation}}
\newcommand\ee{\end{equation}}

\begin{document}

\vspace{5mm}
\vspace{0.5cm}
\begin{center}

\def\thefootnote{\fnsymbol{footnote}}

{\Large \bf Black Hole Superradiance in $f(R)$ Gravities }
	\\[0.7cm]

Mohsen Khodadi$^1$ \footnote{m.khodadi@ipm.ir},
Alireza Talebian$^1$ \footnote{talebian@ipm.ir},
Hassan Firouzjahi$^{1, 2}$ \footnote{firouz@ipm.ir}
\\[0.5cm]

 {\small \textit{$^1$School of Astronomy, Institute for Research in Fundamental Sciences (IPM), \\ P.~O.~Box 19395-5531, Tehran, Iran
}}\\

{\small \textit{$^2$ Department of Physics, Faculty of Basic Sciences, University of Mazandaran, \\
P. O. Box 47416-95447, Babolsar, Iran}}\\

\date{\today}

\end{center}

\vspace{.8cm}

\hrule \vspace{0.3cm}

\begin{abstract}
We study the superradiance amplification factor (SAF) for
a charged massive scalar wave scattering off small and slowly rotating Kerr-Newman black holes in $f(R)$ gravity immersed in asymptotically  flat and de-Sitter spacetimes.  We employ the \emph{``analytical asymptotic matching''} approximation technique which is  valid
in low frequency regime where the Compton wavelength of the propagating particle is much larger than the size of the black hole.
The $f(R)$-Kerr-Newman family solution induces an extra  distinguishable effect on the contribution of the  black hole's electric charge to the metric and that in turn affects the SAFs and their frequency ranges.  While our analysis are general, we present the numerical results for the Starobinsky and Hu-Sawicki $f(R)$ models of gravity as our working examples.
In the case of asymptotically flat spacetime, the SAFs predicted in Starobinsky $f(R)$ model are not distinguishable from those of GR while  for Hu-Sawicki model the SAFs can be weaker or stronger
than those of GR within the frequency parameters space. In the case of  asymptotically de-Sitter spacetime, the superradiance scattering may not either occur in Starobinsky model  or has a weaker chance compared to GR while in Hu-Sawicki  model the results of SAFs and their frequency regimes are different from the standard ones.
 \end{abstract}
\vspace{0.5cm} \hrule
\def\thefootnote{\arabic{footnote}}
\setcounter{footnote}{0}
\newpage
\section{Introduction}

In systems with the capability to dissipate energy there is the possibility of superradiance in which
radiation is enhanced. This phenomenon occurs
in various branches of physics such as in quantum mechanics \cite{QM} and relativity \cite{Zed},
see \cite{Arderucio:2014oua} for a  review. One useful setup to study the superradiance is to look for the scattering of  scalar fields by certain systems in which the scattered field obtains a larger energy compared to the incident field. Black holes are the  favourite candidates for the superradiance to occur
since the event horizon (EH) provides a dissipative mechanism \cite{Brito:2015oca}\footnote{Recall that in context of curved spacetime physics there is also another energy
extraction phenomenon known as the \emph{``Penrose process'' } \cite{Penrose1971a,Penrose1971b} which commonly
thought to be as particle analog of superradiance. Even though, the nature of these two energy extraction processes is generally distinct \cite{Brito:2015oca}, however under some circumstances
one can find some interesting connections between them, see \cite{Vicente:2018mxl}.}.  
In  a composed system of black hole and the external field, superradiance
is equivalent  to the energy extraction of vacuum by the superradiant scattering.
This phenomenon is more interesting in the light of  the detection of the gravitational wave signal such as in \textit{``GW150914'' } (originated by the binary black hole merger) or from the \textit{``Event Horizon Telescope'' } (EHT) \cite{Akiyama:2019cqa} which demonstrates the reality of black holes in nature.

Historically, the study of black hole superradiance stem from the seminal  works of Zeldovich \cite{Zed}
and Misner \cite{Misner:1972kx}  who  predicted the possibility of amplification
of some waves by Kerr black holes. Teukolsky \cite{Teukolsky:1972my} has presented the master equation for the Kerr geometry from the linearized bosonic perturbations (scalar, electromagnetic and gravitational)
which turns out to be separable.  Indeed, by using this master equation for each of scalar, electromagnetic and gravitational waves scattering off a Kerr black hole, Teukolsky and Press were able to
show  that there are some superradiant modes \cite{Teukolsky:1974yv}. While there is no mathematical proof for the absence of superradiance for the fermionic fields, but  Unruh \cite{Unruh:1973bda} and Chandrasekhar \cite{Chan}
demonstrated this conclusion  for the massless and massive Dirac fields scattered by a Kerr black hole respectively.  Bekenstein \cite{Bekenstein:1973mi}, by discovering the relation between the superradiance and Hawking's area theorem, was able to show that
this phenomena can be understood through the classical laws of black hole mechanics.

Superradiance phenomena is not restricted to  black holes arising
from general relativity (GR) but it would happen in any extended theory of gravity that admits black hole solutions. To have an analytic study of superradiance amplification for Kerr-like black holes in extended theories of gravity we have to work in slow-rotation limit \cite{Pani:2011gy} while  going beyond this limit requires  numerical analysis \cite{Kleihaus:2011tg,Delsate:2018ome,Cunha:2019dwb}. One important motivation for studying superradiance in  black hole solutions of modified gravity is that
the geometric structure of black holes indeed contains some information about the modified theory of gravity in the strong field limit. For instance, it is shown in \cite{Pani:2011gy} that for the slowly-rotating black hole solutions predicted by quadratic gravity the proper volume
of the ergoregion decreases. Namely, the background geometry  causes the weakening of the superradiant amplification factor. This means that the superradiance phenomena is sensitive to the geometric structure of black holes so that one may use this phenomenon as a tool to shed light on gravity in the strong field limit.
The analysis of  \cite{Cardoso:2013opa, Cardoso:2013fwa} for the Kerr black holes in scalar-tensor theories
show that the underlying phenomena is sensitive to the presence of matter too.
An interesting  issue  following the superradiance phenomena  in the context of alternative theories
of gravity is the stability analysis of the superradiant modes \cite{Zhang:2014kna}-\cite{Kolyvaris:2018zxl}. Of course, in the context of standard GR  numerous studies have been performed
with a variety of assumptions on the background geometry as well as the field
perturbations (see e.g. \cite{Cardoso:2004hs}-\cite{Benone:2019all} and references therein). Note that the Kerr superradiant instability arising from the hypothetical ultralight bosons such as axions, as one of the candidates for dark matter, have interesting theoretical as well as observational implications.
For the case of real bosonic fields, the cloud disperses for a long time so, depending on the boson masses, it is expected to generate the gravitational wave signals in specific range of
frequencies \cite{Arvanitaki:2016qwi}. However, for the case of  complex bosonic fields, the gravitational wave emission  is suppressed so at the final state of instability a composite system of Kerr black hole
plus an external bosonic structure remains \cite{Herdeiro:2014goa}. This phenomenon can be used to test some  fundamental paradigms in theoretical physics such as the no-hair conjuncture \cite{Cunha:2019ikd}.
Finally,  due to  the spin down instability i.e. the transfer of energy and angular momentum from Kerr black hole to the bosonic cloud, it is possible to impose some constraints on the boson fields
\cite{Cardoso:2018tly}.

With these discussions in mind in this work we would like to address the natural question that what is the effect of  curvature corrections on the superradiance phenomenon?  We would like to address the consequence of curvature corrections on the scalar wave amplification and whether or not
 the deviations from standard GR affect the black hole superradiance.  For this purpose,                                                                                                                                                                                                                                                 we focus on a natural extension of GR, the so called $f(R)$ gravity. From theoretical standpoint, one
advantage of  $f(R)$ gravity compared to  other theories of modified gravity
 is the absence of ghost instabilities \cite{Sotiriou:2008rp, DeFelice:2010aj}. The cosmological and astrophysical implications of $f(R)$ gravity scenarios have been studied extensively, see for example
 \cite{Capozziello:2011et, Nojiri:2010wj, Nojiri:2017ncd, delaCruzDombriz:2008cp} in addition \cite{Sotiriou:2008rp, DeFelice:2010aj}. Due to importance and far reaching implications of $f(R)$ theories in cosmology and astrophysics, it is well motivated to study  the  black hole superradiance in $f(R)$ gravity.  For this purpose, we study the superradiance for the most general black hole solution including the black hole rotation
and charge (Kerr-Newman type solution). This type of black hole solution allows us to study the non-linear interplay between gravity and electromagnetism.
Contrary to the usual belief that the real black holes in sky are mostly electrically neutral, however, some processes in both classical and relativistic frameworks indicate a small non-zero charge for the black holes \cite{Zajacek:2019kla}. Theoretically, there are some mechanisms such as the imbalance between the mass of protons and electrons within the ionized plasma around the black hole and or the twisting of magnetic field lines due to rotation which allow the black hole to be charged, see \cite{Zajacek:2018ycb}.

While our analysis are for general $f(R)$ theories, we present the numerical results for the two most interesting examples of $f(R)$ gravity: the Starobinsky model \cite{Starobinsky:1980te} and the Hu-Sawicki model \cite{Hu:2007nk}. The Starobinsky model is the first inflationary model which is well consistent with the cosmic microwave background data such as  the Planck observations \cite{Ade:2015xua, Ade:2015lrj}. The Hu-Sawicki model, on the other hand, may be viewed  as a counterpart to $\varLambda CDM$  while at the same time satisfying the standard tests of the solar system via  a mechanism known as the ``chameleon screening''.  In order to provide a realistic study, our discussion will cover both asymptotically flat and de-Sitter $f(R)$ Kerr-Newman black hole spacetimes.

The rest of the  paper is organized as follows. After an overview of Kerr-Newman black hole solutions
in $f(R)$ theory in Sec. \ref{Br}, we determine the relevant superradiance conditions for
asymptotically flat and de-Sitter Kerr-Newman black hole spacetimes in Sec. \ref{C}.  In Sec. \ref{Su} the analytic expressions for the superradiance amplification factor (SAF) of  $f(R)$ Kerr-Newman black hole and charged massive scalar field  are presented with Starobinsky and Hu-Sawicki models as our case studies. The summary and discussions are presented in Sec. \ref{Sum}.  We work in natural unites $c=\hbar=k_B=G_N=1$.

\section{Kerr-Newman Black Hole in  $f(R)$ Gravity}\label{Br}

We study the charged black hole solutions in $f(R)$ modified gravity
which are either asymptotically flat or are in dS space, so we assume the spacetime has a constant curvature $R=R_0$. The action of the system is
\begin{eqnarray}\label{sf}
S = \int \text{d}^{4}x\sqrt{\mid g\mid}\,\bigg(f(R)+2\kappa^2 {\cal L}_{\rm em}\bigg)\,, ~~~~~~~~~~~~\kappa^2 = 8\pi\,,
\end{eqnarray}
in which $R$ is the Ricci scalar and $g$ denotes the determinant
of the metric $g_{\mu\nu}$. There is an electric field which has filled the spacetime with the Lagrangian density  ${\cal L}_{\rm em}=-{1 \over 4}F_{\mu\nu}F^{\mu\nu}$ in which $A_\mu$ is the vector potential and
$F_{\mu\nu}=\partial_\mu\,A_\nu-\partial_{\nu}A_\mu$
is the field strength tensor obeying the Maxwell equation $\partial_\mu\,(\sqrt{-g}F^{\mu\nu})=0$.

Varying action \eqref{sf} with respect to the inverse metric, we obtain the modified Einstein equations
\begin{eqnarray}
 R_{\mu\nu}\,f'(R_0)-\frac{1}{2}\,g_{\mu\nu}\,f(R_0)
= 8\pi\, T_{\mu\nu}\,,
\label{ec_tensorial}
\end{eqnarray} where  $T_{\mu\nu}=F_{\mu\alpha}F_{\,\,\,\nu}^\alpha
-\frac{1}{4}g_{\mu\nu}F_{\alpha\beta}
F^{\alpha\beta}$ is the stress-energy tensor of the electromagnetic field.
Taking the trace of Eq. (\ref{ec_tensorial}) in the absence of matter sources, one obtains
the constant curvature scalar \cite{Cognola:2005de}
\begin{eqnarray}
 R_0=\frac{2f(R_0)}{f'(R_0)}\equiv 4\Lambda_f \,,
\label{ec_escalar}
\end{eqnarray}
where $\Lambda_f$ is the cosmological constant  associated with  the curvature constant $R_0$ so the
cases $R_0=0$, $R_0>0$ and $R_0<0$, corresponds to the flat, de-Sitter and anti de-Sitter spacetimes, respectively. Having defined $\Lambda_f$,  Eq. (\ref{ec_tensorial}) can now be rewritten as
\begin{eqnarray}
R_{\mu\nu}=\Lambda_f g_{\mu\nu}+\frac{8\pi}{f'(R_0)} T_{\mu\nu}.
\label{ec}
\end{eqnarray}

Adopting the standard Boyer-Lindquist coordinate $(t,r,\vartheta,\phi)$, the four-dimensional
axisymmetric and stationary solution in $f(R)$ gravity with a constant curvature scalar $R_0$ is given by \cite{Cembranos:2011sr, delaCruzDombriz:2010xy, delaCruzDombriz:2009et, Nojiri:2017kex, Nojiri:2014jqa}
\begin{align}
\mathrm{d}s^2 &=g_{\alpha\beta}dx^\alpha dx^\beta \\&=-\frac{\Delta_r}{\rho^2\chi^2 }\left(\mathrm{d}t-a\sin^2\vartheta
\mathrm{d}\phi\right)^2+\frac{\rho^2}{\Delta_r}\mathrm{d}r^2+
\frac{\rho^2}{\Delta_\vartheta}\mathrm{d}\vartheta^2
+\frac{\Delta_\vartheta
 \sin^2\vartheta}{\rho^2\chi^2}
\left(a~\mathrm{d}t-(r^2+a^2)\mathrm{d}\phi\right)^2\;,\nonumber\\[2ex]
&~~~~~~~~~~~~~~~~~t\in(-\infty,\infty),~~~ r\in(0,\infty),~~~\vartheta\in[0,\pi],~~~ \phi\in[0,2\pi) \;,
\label{kerr}
\end{align}
with
\begin{align}
\rho^2& \equiv r^2+a^2\cos^2\vartheta\;,\nonumber\\
\Delta_r& \equiv \left(r^2+a^2\right)\left(1-\frac{R_0}{12}\,r^2\right)-2Mr+\frac{q^2}{f'(R_0)} \;,\nonumber \\[2ex]
\Delta_\vartheta& \equiv 1+\frac{R_0}{12}\,a^2\cos^2\vartheta\;,\nonumber\\[2ex]
\chi& \equiv 1+\frac{R_0}{12}\,a^2\;.
\label{definitions}
\end{align}
With the above metric, the potential vector as well as the electromagnetic field tensor required in Eq. (\ref{ec_tensorial})
take the following forms
 \begin{eqnarray}\label{vec}
&A_{\alpha}&=\frac{q\,r}{\chi\,\rho^2}\big(-1,0,0,a\,\sin^2
{\vartheta}\big) \, , 
\end{eqnarray}
and
 \begin{eqnarray}
&F_{\alpha\beta}&=\Bigg(
                                                 \begin{array}{cccc}
                                                   0 & \frac{q(r^2-a^2\cos^2\vartheta)}{\rho^4 \chi}& \frac{q r a^2\sin2\vartheta}{\rho^4 \chi} & 0 \\
                                                   -\frac{q(r^2-a^2\cos^2\vartheta)}{\rho^4 \chi}& 0 & 0 & \frac{aq\sin^2\vartheta(r^2-a^2\cos^2\vartheta)}{\rho^4 \chi} \\
                                                    -\frac{q r a^2\sin2\vartheta}{\rho^4 \chi}& 0 & 0 & -\frac{qra^2(r^2+a^2)\sin2\vartheta}{\rho^4 \chi} \\
                                                   0 & -\frac{aq\sin^2\vartheta(r^2-a^2\cos^2\vartheta)}{\rho^4 \chi} & \frac{q ra^2(r^2+a^2)\sin2\vartheta}{\rho^4 \chi}& 0 \\
                                                 \end{array}
                                               \Bigg) \, .
\label{EM}
\end{eqnarray}
A distant  observer may interpret the above solution as a  Kerr-Newman family of black hole with mass $M$, the angular momentum per unite mass  $a\equiv J/M$ and the electric charge $q$.

Compared to  the case of  GR, here the contribution of the black hole's electrical charge to the metric is modified by the factor  ${f'(R_0)}^{-1/2}$  as seen from the definition of $\Delta_r$.   For simplicity,
from now on we use the notion $Q\equiv\frac{q}{\sqrt{f'(R_0)}}$. However, note that the electrical charge of the black hole as measured  by the  distant observer is $q$ and not $Q$,
as is evident from the vector potential and the field strength in Eqs. (\ref{vec}) and (\ref{EM}), respectively.

Alternatively, one can look at the effect of $f(R)$ as follows. By restoring  the gravitational constant $G_N$ then the correction arising from $f(R)$ may be viewed as  an effective gravitational constant $G_{eff}=\frac{G_N}{f'(R_0)}$ instead of effective charge $Q$. In this way, the definition of $\Delta_r$ above is re-expressed as $\Delta_r \equiv \left(r^2+a^2\right)\left(1-\frac{R_0}{12}\,r^2\right)-2G_{N} M r+G_{eff}q^2$ which, after  fixing $G_N=1$, is equivalent to $\Delta_r$ in Eq. (\ref{definitions}). Therefore,  the imprints of curvature correction can be captured either by an effective charge or by an effective gravitational constant.

Defining the horizon via  $g^{rr}=\Delta_r=0$,  we obtain the following quartic equation
\begin{eqnarray}
R_0 r^4+(R_0 a^2-12)r^2+24M r-12(a^2+Q)=0 \, ,
\end{eqnarray}
which yields  four roots, $r_{1}, ..., r_{4}$.

For the flat spacetime ($R_0=0$) the above equation has two positive real roots: $r_{c,h}=M\mp\sqrt{M^2-(a^2+Q^2)}$, representing the positions of the Cauchy and the event horizons receptively.  However, if $R_0>0$, we have three positive roots $r_{c,h}$ and $r_{H}$ in which $r_{H}$ represents the cosmological horizon.  For the  case of $R_0<0$ there are just two positive roots (as in the case of flat spacetime).

For convenience, we define the following  parameters,
\begin{eqnarray}\label{Ang}
\Omega_h \equiv \dfrac{a}{r_{h}^2+a^2}\;,~~~~\Omega_{H} \equiv \dfrac{a}{r_{H}^2+a^2}\;,~~~~ \Phi_{h}\equiv~\frac{Q\,
r_h}{(r^2_h+a^2)}\, ,
\end{eqnarray}
which respectively represent the angular velocity $\Omega$ on the surfaces of event horizon and cosmological horizon and the electric potential $\Phi$ on the surface of event horizon.

\section{Condition for Superradiance Modes}\label{C}

To study superradiance, we consider a complex scalar field $\Psi$ with mass $\mu_s$ which is charged under
the $U(1)$ gauge field with the electric charge coupling \textbf{e}. The corresponding Klein-Gordon equation is
\begin{align}
(\Box-\mu^2_s)\Psi=\frac{1}{\sqrt{\mid g\mid}}
D_\alpha \big(\sqrt{\mid g\mid}g^{\alpha \beta}D_\beta\Psi \big)-\mu^2_s\Psi=0\;,
\label{KG}
\end{align}
where the covariant derivative is given by
$D_\alpha=\partial_\alpha-i\textbf{e}A_\alpha$.

To solve the Klein-Gordon equation, we introduce the following ansatz
 \begin{eqnarray}
\Psi=e^{-i\omega t+im\phi}\mathcal{R}(r)S(\vartheta)\;,
\end{eqnarray}
with the positive oscillation frequency $\omega>0$ and the azimuth angular number $m$.
Inserting the above ansatz into (\ref{KG}) we obtain the following separated differential
equations for $\mathcal{R}(r)$ and $S(\vartheta)$
\begin{eqnarray}
\Delta_r \frac{d}{dr}\Big(\Delta_r\frac{d}{dr} \mathcal{R}(r)\Big)+
\Big[\chi^2 \big(\omega(r^2+a^2)-ma- \textbf{e} q r\big)^2-\Delta_r(\lambda+\mu^2_s~r^2+a^2\omega^2-2ma\omega)
\Big]\mathcal{R}(r)=0\;,
\label{KGR}
\end{eqnarray}
and
\begin{eqnarray}
\sin\vartheta\frac{d}{d\vartheta}\bigg(\sin\vartheta\frac{d}
{d\vartheta} S(\vartheta)\bigg)
+\bigg(\lambda\sin^2\vartheta
+\frac{a^2(\omega^2-\mu_s^2)}{4} \sin^2 2\vartheta-m^2\bigg)S(\vartheta)=0\, .
\end{eqnarray}
Here $\lambda \equiv l (l+1)$ denotes the angular separation constant with non-negative
angular momentum index $l\ge 0$.
From now on, we define $\varepsilon\equiv\textbf{e}q$  representing the joint coupling of the scalar field and the black hole electric charges. Increasing the value
of $\varepsilon$, this coupling becomes large and we enter the strong coupling limit when
$\varepsilon>1$ \footnote{This regime seems to be relevant for the
expected black holes in our universe with even small charges \cite{Hod:2018dpx}.}.

Defining the new field variable $u(r)\equiv \sqrt{r^2+a^2}~\mathcal{R}(r)$ and going to the tortoise coordinate
defined via $dr_*=\dfrac{r^2+a^2}{\Delta_r}\;dr$, after some algebra Eq. (\ref{KGR}) takes the following Schrodinger-like form
\begin{eqnarray}
 \frac{d^2u(r_*)}{dr^{*2}}+V_{eff}(r)\,u(r_*)=0\;,
\label{Sch}
\end{eqnarray}
with the effective potential given by
\begin{eqnarray}
V_{eff}=&&\chi^2\Big[\omega-\frac{m a}{r^2+a^2}-\frac{\varepsilon \, r}{(r^2+a^2)}\Big]^2-\frac{\Delta_r}{(r^2+a^2)^2}\times \nonumber\\
&&\left[
	\lambda+\mu_s^2 r^2+a^2\omega^2-2ma\omega+ \sqrt{r^2+a^2}\frac{d}{dr}\big(\frac{r \Delta_r}{(r^2+a^2)^{3/2}}\big)\right].
\label{Potential}
\end{eqnarray}

Now we consider the asymptotic behaviour of the solutions for the flat and de-Sitter backgrounds separately. For the
 flat background $R_0=0$, the asymptotic solutions of Eq. (\ref{Sch}) reads off as
\begin{align}\label{Asy-so}
 u_{h}(r) =& \mathcal{A_{T}} \exp(-ik_h r_*), \qquad  \qquad \qquad \qquad \qquad r_*\longrightarrow-\infty~ (r \rightarrow r_{h}) ,\nonumber\\
u_\infty(r)=& \mathcal{A_{I}}~r^{b}~\exp(-ik_\infty r_*)+\mathcal{A_{R}}~r^{b}~\exp(ik_\infty r_*),    \qquad r_*\longrightarrow\infty~ (r \rightarrow \infty),
\end{align}
where $k_h=\sqrt{V_{eff}(r\rightarrow r_h)}=\left(\omega-(m +\frac{\varepsilon r_{h}}{a})\Omega_{h}\right)$, $k_\infty=\sqrt{V_{eff}(r\rightarrow \infty)}=\sqrt{\omega^2-\mu_s^2}$ and $b\equiv i\frac{M\mu^2-\varepsilon\omega}{k_\infty}$.

Similarly,  for the de-Sitter spacetime ($R_0>0$), we have
\begin{align}\label{ds-so}
u_{h}(r)=& \mathcal{A_{T}} \exp(-ik_h r_*), \qquad  \qquad \qquad \qquad \qquad r_*\longrightarrow-\infty~ (r \rightarrow r_{h}) ,\nonumber\\
u_{H}(r)=& \mathcal{A_{I}}~ \exp(-ik_{H} r_*)+\mathcal{A_{R}}~\exp(ik_{H} r_*), \qquad  r_*\longrightarrow\infty~ (r \rightarrow r_{H}),
\end{align}
where here $k_h=\chi\left(\omega-(m +\frac{\varepsilon r_{h}}{a})\Omega_{h}\right)$ and $k_{H}=\sqrt{V_{eff}(r\rightarrow r_{H})}=\chi\left(\omega-(m +\frac{\varepsilon r_{H}}{a})\Omega_{H}\right)$.

The boundary condition (\ref{Asy-so}) represents an incoming wave with the
amplitude $\cal{A_{I}}$ which comes from spatial infinity so that after
scattering off the event horizon it gives rise to a reflected and transferred
waves with the amplitudes $\cal{A_{R}}$ and $\cal{A_{T}}$ respectively.
However, the boundary condition (\ref{ds-so}) tell us the incoming wave
originates from the cosmological horizon and after being scattered off the black hole,
it gives rise to a reflected wave which goes back to the cosmological horizon
and a transferred wave which passes through the black hole's event horizon.

Now, by equating the Wronskian for regions near the event horizon $W_h=(u_{h}\frac{d u^*_{h}}
{dr_*}-u^*_{h}\frac{d u_{h}}{dr_*})$ with its other counterparts at infinity and
on cosmological horizon $W_{\infty(H)}=(u_{\infty (H)}\frac{d u^*_{\infty (H)}}{dr_*}-u^*_{\infty (H)}\frac{d u_{\infty (H)}}{dr_*})$,
we arrive at the following conditions
\begin{eqnarray}\label{flat}
\mathcal{|A_{I}|}^2-\mathcal{|A_{R}|}^2=\frac{\omega-(m +\frac{\varepsilon r_{h}}{a})\Omega_{h}}
{\sqrt{\omega^2-\mu_s^2}}\mathcal{|A_{T}|}^2\,, \qquad \qquad (\mathrm{flat ~ background})
\end{eqnarray}
and
\begin{eqnarray}\label{ds}
\mathcal{|A_{I}|}^2-\mathcal{|A_{R}|}^2=\frac{\omega-(m +\frac{\varepsilon r_{h}}{a})\Omega_{h}}
{\omega-(m +\frac{\varepsilon r_{H}}{a})\Omega_{H}}\mathcal{|A_{T}|}^2\,, \qquad \qquad (\mathrm{dS ~ background})
\end{eqnarray}
for the flat and de-Sitter spacetimes, respectively.

In order for the superradiance to take place the amplitude of the reflected wave must exceed the amplitude of the incident wave so the following frequency
conditions must be met
\begin{eqnarray}\label{Sflat}
\mu_s<\omega<(m +\frac{\varepsilon r_{h}}{a})\Omega_{h}~, ~~~~~~ \qquad \qquad (\mathrm{flat ~ background})
\end{eqnarray}
and
\begin{eqnarray}\label{Sds}
(m +\frac{\varepsilon r_{H}}{a})\Omega_{H}<\omega<(m +\frac{\varepsilon r_{h}}{a})\Omega_{h}~, \quad  (\mathrm{dS ~ background})
\end{eqnarray}
for flat and de-Sitter backgrounds respectively.

At first glance, however, one might imagine that modifications in the frequency conditions  (\ref{Sflat}) and (\ref{Sds}) are just a renormalization of the black hole's electric charge $q$ or $\varepsilon$. Although mathematically it seems to be true, physically this is not the case. In fact, the physical electrical charge of the black hole as measured by a distant observer is still $q$ ( as we have already addressed through Eqs. (\ref{vec}) and (\ref{EM})), meaning that the correction induced by $f(R)$ modified gravity on the black hole's charge are distinct from each other. So, in essence in Eqs. (\ref{Sflat}) and (\ref{Sds}), we deal with a
new  distinguishable contribution which comes directly from gravitational corrections. The aforementioned equations indicates that the  $f(R)$ correction affects the superradiance conditions
compared to GR (with $f'(R_0)=1$). More specifically, in the presence of $f(R)$ correction with $f'(R_0)\neq1$,  the threshold
superradiance frequency, $\omega_t \equiv (m +\frac{\varepsilon r_{h}}{a})\Omega_{h}$, is modified  relative to
its GR counterpart. Since the onset of  superradiance instability in the composed system consisting of Kerr-Newman black hole and the massive scalar field is characterized by $\omega=\omega_t$,  the displacement in the threshold frequency can be of phenomenological importance. In the case of instability
occurring  due to superradiance scattering,\footnote{Note that superradiance scattering does not always create instability in the system under question. For instance, it is shown in \cite{Hod:2013eea}
that the superradiance scattering of charged massive scalar field does not lead to instability in Reissner-Nordstrom black hole when $Q/M\leq2\sqrt{2}/3$.} the threshold frequency in essence is a boundary with marginal stability, separating stable ($\omega>\omega_t$) and unstable regions.
With these discussions in mind , in next section, we investigate the effects of curvature modifications on the range of superradiance frequency as well as the power of superradiance for both asymptotically flat and de-Sitter spacetimes.

Before proceeding, however, let us here mention an interesting point.
If we take the limit $R_0\rightarrow0$ in Eq. (\ref{Sds}), its lower bound does not coincide with Eq. (\ref{Sflat}) since $\Omega_H$ goes to zero as $R_0\rightarrow0$. This mismatch was already seen in Kerr-de-Sitter black holes \cite{Tachizawa:1992ue} where the authors have argued that, despite the oddity of this difference, there seems to be something else going on. Inspired from Ref. \cite{Tachizawa:1992ue} one can conclude  that when $R_0>0$, superradiance always occurs if Eq. (\ref{Sds}) is satisfied.
However, for $\omega<\mu_s$ the tunnelling probability (proportional to $\mathcal{|A_{T}|}^2$) becomes much smaller than that for $ \mu_s<\omega<(m +\frac{ \varepsilon r_{H}}{a})\Omega_{H}$ so the superradiance amplification is extremely suppressed. So, as $R_0\rightarrow 0$, the superradiance amplification vanishes for waves in the range $(m +\frac{\varepsilon r_{H}}{a})\Omega_{H}<\omega<\mu_s$, which is consistent with condition (\ref{Sflat}).

\section{Superradiance Amplification Factors}\label{Su}

Despite the fact that  the Teukolsky's equation (in particular the radial equation (\ref{KGR}))
can not be solved analytically, some approximate
methods have been developed. In this section, using the \textit{``analytical asymptotic matching''} (AAM)
method\footnote{The AAM method is indeed a common approach in finding an accurate
approximation solution for a singularly perturbed differential equation. In other word,
if the exact solution is not available we may still be able to construct an approximate
solution using the inner and outer asymptotic expansions. The principle idea of this method is to
find different approximate solutions where each one is valid for part of the range of
the independent variable. Combining them, one arrives at a single approximate solution
 for the original equation.}, proposed first by Starobinsky
\cite{Starobinsky:1973aij}, we obtain the ``amplification factor'' of a scalar wave scattering off a $f(R)$-charged
Kerr black hole. This enables us to detect the effect of $f(R)$ correction
on the amplification factor $Z_{lm}\equiv \frac{\mathcal{|A_{R}|}^2}{\mathcal{|A_{I}|}^2}-1$,
a dimensionless quantity which its positive value indicates a
superradiant amplification from the black hole. To employ the AAM method we have to impose
the approximation that the Compton wavelength of the propagating particle
is very large compared to the size of the black hole, i.e. $\mu_s r_h\ll1$. In addition,
the slow rotation approximation (i.e. $a\omega\ll1$) is usually employed in this method \cite{Detweiler:1980uk}.
However,  there are other approaches
such as the partial wave method which does not require the slow rotation approximation \cite{Benone:2019all}.
\\

The main point in employing AAM method is that one can split the space
outside the event horizon into two limits: region near the  horizon
($r-r_h\ll \omega^{-1}$) known as the \emph{``near-region''}, and region
very far from the horizon
($r-r_h\gg M$) known as the \emph{``far-region''}. The exact
solutions derived for the above two asymptotic regions are matched
 in an overlapping region where $M\ll r-r_h\ll \omega^{-1}$.
However, this method has two obvious limitations. First, to applying it
the parameters involved in the equation must obey some certain conditions.
Here, one requires  that  $M\omega\ll1$, $\mu_sM\ll1$ and $\varepsilon\ll1$.
Therefore, in order to apply the AAM method, our analysis is restricted to some certain frequency
parameter space along with the assumption of the weak coupling between charged scalar field and Kerr-Newman black hole.
Second, matching is possible only when the relevant expansions have overlaping regions.
So, as a further limitation, the approximation becomes less reliable as one deviates  from the overlaping region i.e. $\mu_sM\ll r-r_h\ll1$.
Indeed, when $r-r_h$ approaches to the extremal points  $\mu_sM$ ($M\omega$) or $1$,
then the error in approximate solution becomes significant and the solution may not be trusted.
In the following, to provide an analytic expression for the amplification factors
of a scalar wave scattering off a $f(R)$-charged Kerr black
hole, we solve the radial equation \eqref{KGR} using the above approximations.

\subsection{Asymptotically flat spacetime: $R_0=0$}

In this subsection we preset  the superradiance analysis for a black hole located in an
asymptotically flat spacetime, $R_0=0$.

\paragraph{(i) Near-region solution:}

First we obtain the solution for the near-region.

Performing the change of variable $x=\frac{r-r_h}{r_h-r_c}$ and
plugging $\triangle_r \frac{d}{dr}=(r_h-r_c)x(x+1) \frac{d}{dx}$
into Eq. \eqref{KGR}, we obtain
\begin{eqnarray}
\label{nr}
x^{2}(x+1)^2\frac{\mathrm{d}^2\mathcal{R}}{\mathrm{d}x^2}&+&x(x+1)(2x+1)
\frac{\mathrm{d}\mathcal{R}}{\mathrm{d}x}  \nonumber\\
&+&\left[\eta^2-l(l+1) x(x+1)-
\mu_s^2\left((r_h-r_c) x+r_h\right)^2x(x+1)\right] \mathcal{R}=0\,,
\end{eqnarray}
where  we have defined $\eta\equiv\frac{r_h^2+a^2}{r_h-r_c}
\big(\omega-(m +\frac{\varepsilon r_{h}}{a})\Omega_{h}\big)$.

For regions near the horizon, we can approximate $\mu_s^2\left((r_h-r_c) x+r_h\right)^2
\approx \mu_s^2r_h^2$ and further by applying the Compton wavelength
approximation, the above equation simplifies to
\begin{eqnarray}\label{nrr}
x^{2}(x+1)^2\frac{\mathrm{d}^2\mathcal{R}}{\mathrm{d}x^2}+x(x+1)(2x+1)\frac{\mathrm{d}\mathcal{R}}{\mathrm{d}x}
+\left(\eta^2-l(l+1) x(x+1)\right)\mathcal{R}=0\,.
\end{eqnarray}
The most general solution of Eq.~\eqref{nrr} is given in terms of
ordinary hypergeometric functions $_2F_1(a,b;c;z)$
\begin{eqnarray}\label{general}
\mathcal{R}(x) &=& C_1~x^{-i\eta}(1+x)^{-i\eta}~_2F_1(-l-2i\eta,l+1-2i\eta;1-2i\eta;-x)
\nonumber\\
&+& C_2~x^{i\eta}(1+x)^{-i\eta}~_2F_1(-l,l+1;1+2i\eta;-x) \,.
\end{eqnarray}
Imposing the ingoing boundary condition,
and also using the following identities
\begin{eqnarray}
_2F_1(a,b;c;z) &=& z^{1-c}(1-z)^{c-a-b}~_2F_1(1-a,1-b;2-c;z)\,, \\
_2F_1(a,b;c;z) &=& _2F_1(b,a;c;z) \,,
\end{eqnarray}
the solution \eqref{general} finally reads off
\begin{eqnarray}
\mathcal{R}_{\mathrm{near}}=C~\big(\frac{x}{x+1} \big)^{i\eta}~_2F_1(-l,l+1;1-2i\eta;-x) \,.
\end{eqnarray}

To match the above solution to the solution from the  far-region, we consider the behaviour of above solution at large $x$, yielding
\begin{eqnarray}
\label{solnear}
\mathcal{R}_{\mathrm{near- large~x}}\sim C~\Big(\frac{\Gamma(2l+1)\Gamma(1-2i\eta)}{\Gamma(l+1-2i\eta)\Gamma(l+1)}~x^{l}+
\frac{\Gamma(-2l-1)\Gamma(1-2i\eta)}{\Gamma(-l)\Gamma(-l-2i\eta)}x^{-l-1} \Big)\, ,
\end{eqnarray}
where the approximation $(x+1)^{i\eta}\approx x^{i\eta}$ along with the
following asymptotic behaviour of the hypergeometric function has been used
\begin{eqnarray}
\lim\limits_{x \rightarrow \infty}~ _2F_1(a,b;c;-x) = \dfrac{\Gamma(b-a)\Gamma(c)}{\Gamma(c-a)\Gamma(b)}x^{-a}+\dfrac{\Gamma(a-b)\Gamma(c)}{\Gamma(c-b)\Gamma(a)}x^{-b}\, .
\end{eqnarray}

\paragraph{(ii) Far-region solution:}
Now we consider the solution for the far-region.  In the asymptotic region, $x \rightarrow \infty \,  (x\gg1~,~r\gg r_h$), Eq.~\eqref{nr} is simplified to
\begin{eqnarray}\label{far}
\frac{\mathrm{d}^2\mathcal{R}}{\mathrm{d}x^2}+\frac{2}{x}\frac{\mathrm{d}\mathcal{R}}{\mathrm{d}x}+
\left(\xi^2-\frac{l(l+1)}{x^2}\right)\mathcal{R}=0\,,
\end{eqnarray} where $\xi \equiv(r_h-r_c)\sqrt{\omega^2-\mu_s^2}$.

The solution of the above equation is given in terms
of the confluent hypergeometric function of the
second kind $U(a,b,z)$ and the generalized Laguerre polynomial $L_n^{(a)}(z)$ as follows:
\begin{eqnarray}\label{solCL}
\mathcal{R} = \exp({-i \xi x})\left(A_1~x^{l}~U(l+1,2l+2,2i\xi x)+A_2~x^{l}~
L_{-l-1}^{(2l+1)}(2i\xi x)\right)\,.
\end{eqnarray}
Using the following expression
\begin{eqnarray}
L_n^{(a)}(z)=\dfrac{(-1)^n}{n!}~U(-n,a+1,z) \,,
\end{eqnarray} and also the identity $U(a,b,z)= z^{1-b}~ U(1+a-b,2-b,z)$, the solution in  Eq.
\eqref{solCL} takes the following form
\begin{eqnarray}\label{sol}
\mathcal{R}_{\mathrm{far}} = \exp({-i \xi x}) \Big(A_1~x^{l}~U(l+1,2l+2,2i\xi x)+A_2~x^{-l-1}~U(-l,-2l,2i\xi x)\Big)\,.
\end{eqnarray}

To match the above far-region solution to the near-region solution, now we consider the small $x$ limit of the above solution.
Using the Taylor expansion $\lim_{z\rightarrow0}U(a,b,z)\approx\frac{\Gamma(1-b)}{\Gamma(1+a-b)}+....$  the approximate form of the above solution at small $x$ is given by
\begin{eqnarray}\label{solfar}
\mathcal{R}_{\mathrm{far-small \,\, x}}\sim  A_1\frac{\Gamma(-2l-1)}{\Gamma(-l)}~x^{l}+
A_2\frac{\Gamma(2l+1)}{\Gamma(l+1)}~x^{-l-1}\,.
\end{eqnarray}

\paragraph{(iii) Amplification factor using matching:}
Having obtained the solutions for the far-region and the near-region and by matching these two asymptotic solutions,  we can compute the scalar wave fluxes at infinity to obtain the amplification
factor.

Equating Eqs. \eqref{solnear} and \eqref{solfar} we obtain
\begin{eqnarray}
A_1&=& C\frac{\Gamma(-l)\Gamma(2l+1)\Gamma(1-2i\eta)}{\Gamma(l+1-2i\eta)\Gamma(l+1)\Gamma(-2l-1)}\,,\\
A_2&=&C\frac{\Gamma(l+1)\Gamma(-2l-1)\Gamma(1-2i\eta)}{\Gamma(-l)\Gamma(2l+1)\Gamma(-l-2i\eta)}\,.
\end{eqnarray}
In order to compute the scalar wave fluxes at infinity, we have to connect
the coefficients $A_1$ and $A_2$ with coefficients $\mathcal{A_{I}}$ and
$\mathcal{A_{R}}$ in the infinity limit of the radial solution \eqref{Asy-so}.
To do so, we first expand the far region solution \eqref{sol} at infinity as
\begin{align}\label{Asysoo}
A_1 \frac{\Gamma(2l+2)}{\Gamma(l+1)}\xi^{-l-1}\bigg((-2i)^{-l-1}\frac{\exp[-i\xi x]}{x}+
(2i)^{-l-1}\frac{\exp[i\xi x]}{x}\bigg)+\\ \nonumber
A_2 \frac{\Gamma(-2l)}{\Gamma(-l)}\xi^{l}\bigg((-2i)^{l}\frac{\exp[-i\xi x]}{x}+
(2i)^{l}\frac{\exp[i\xi x]}{x}\bigg)\, .
\end{align}
Now by applying the approximations $\frac{1}{x}\sim\frac{\xi}{\sqrt{\omega^2-\mu_s^2}~ r},~~
\exp(\pm i\xi x)\sim\exp(\pm i\sqrt{\omega^2-\mu_s^2} r)$ and then matching the above solution
with the radial solution
\begin{align}\label{Asysooo}
\mathcal {R}_\infty(r)\sim \mathcal{A_{I}}~\frac{\exp(-i\sqrt{\omega^2-\mu_s^2} r^*)}{r}+\mathcal{A_{R}}~
\frac{\exp(i\sqrt{\omega^2-\mu_s^2} r^*)}{r}, \qquad  \qquad
 (r \rightarrow \infty) ,
\end{align} we obtain
\begin{eqnarray}
\mathcal{A_{I}}&=&\frac{1}{\sqrt{\omega^2-\mu_s^2}}\left[A_1 \frac{(-2 i)^{-l-1} \xi^{-l}
\Gamma (2 l+2)}{\Gamma (l+1)}+
A_2\frac{(-2i)^{l} \xi^{l+1} \Gamma (-2 l)}{\Gamma (-l)}\right]\,,\\
\mathcal{A_{R}}&=&\frac{1}{\sqrt{\omega^2-\mu_s^2}} \left[A_1 \frac{ (2 i)^{-l-1}\xi^{-l}
\Gamma (2l+2)}{\Gamma (l+1)}+A_2\frac{ (2i)^{l} \xi^{l+1} \Gamma (-2 l)}{\Gamma (-l)}\right]\,.
\end{eqnarray}
Finally, by substituting the relevant expressions for $A_1$ and $A_2$,
we obtain
\begin{eqnarray}
\mathcal{A_{I}}=\frac{C}{\sqrt{\omega^2-\mu_s^2}}\left[\frac{(-2i)^{-l-1}\xi^{-l}\Gamma(-l)
\Gamma(2l+1)\Gamma(2l+2)(\Gamma(1-2i\eta))^2}
{\Gamma(-2l-1)\big(\Gamma(l+1-2i\eta)\Gamma(l+1)\big)^2}+\right.\nonumber\\\left.
\frac{(-2 i)^{l}  \xi^{l+1} \Gamma (-2l)\Gamma(l+1)\Gamma(-2l-1)\big(\Gamma(1-2i\eta)\big)^2}
{\Gamma(2l+1)\big(\Gamma(-l)\Gamma(-l-2i\eta)\big)^2}\right]\,,
\end{eqnarray}
and
\begin{eqnarray}
\mathcal{A_{R}}=\frac{C}{\sqrt{\omega^2-\mu_s^2}}\left[\frac{(2i)^{-l-1}\xi^{-l}
\Gamma(-l)\Gamma(2l+1)\Gamma(2l+2)(\Gamma(1-2i\eta))^2}
{\Gamma(-2l-1)\big(\Gamma(l+1-2i\eta)\Gamma(l+1)\big)^2}+
\right.\nonumber\\\left.
\frac{(2 i)^{-l-1} \xi^{-l} \Gamma (-2l)\Gamma(l+1)\Gamma(-2l-1)\big(\Gamma(1-2i\eta)\big)^2}
{\Gamma(2l+1)\big(\Gamma(-l)\Gamma(-l-2i\eta)\big)^2}\right]\,.
\end{eqnarray}
As a result, the amplification factor can then be computed via
\begin{eqnarray}\label{amp}
Z_{lm}=\frac{|\mathcal{A_{R}}|^2}{|\mathcal{A_{I}}|^2}-1\,.
\end{eqnarray}

  \begin{table}
    \begin{center}
\begin{tabular}{|c|c|c|c|}
  \hline
   $f(R)$ models& Ref & $f'(0)$ & $f''(0)$ \\
  \hline
  \textbf{Model I:}~ $f=R+\alpha R^2,~~~\alpha>0$ & \cite{Starobinsky:1980te} & 1 & $2\alpha$ \\ \hline
  \textbf{Model II:}~$f^{n=1}=R-\gamma^2\frac{c_1(R/\gamma^2)^n}{1+c_2(R/\gamma^2)^n}$&\cite{Hu:2007nk}&$1-c_1$&$\frac{2c_1c_2}{\gamma^2}$
  \\  \hline
  \end{tabular}
  \caption{Two viable cosmological $f(R)$ models in asymptotically flat spacetime. }
  \label{f(r)}
   \end{center}
 \end{table}
 \begin{figure}[!ht]
	\includegraphics[scale=0.37]{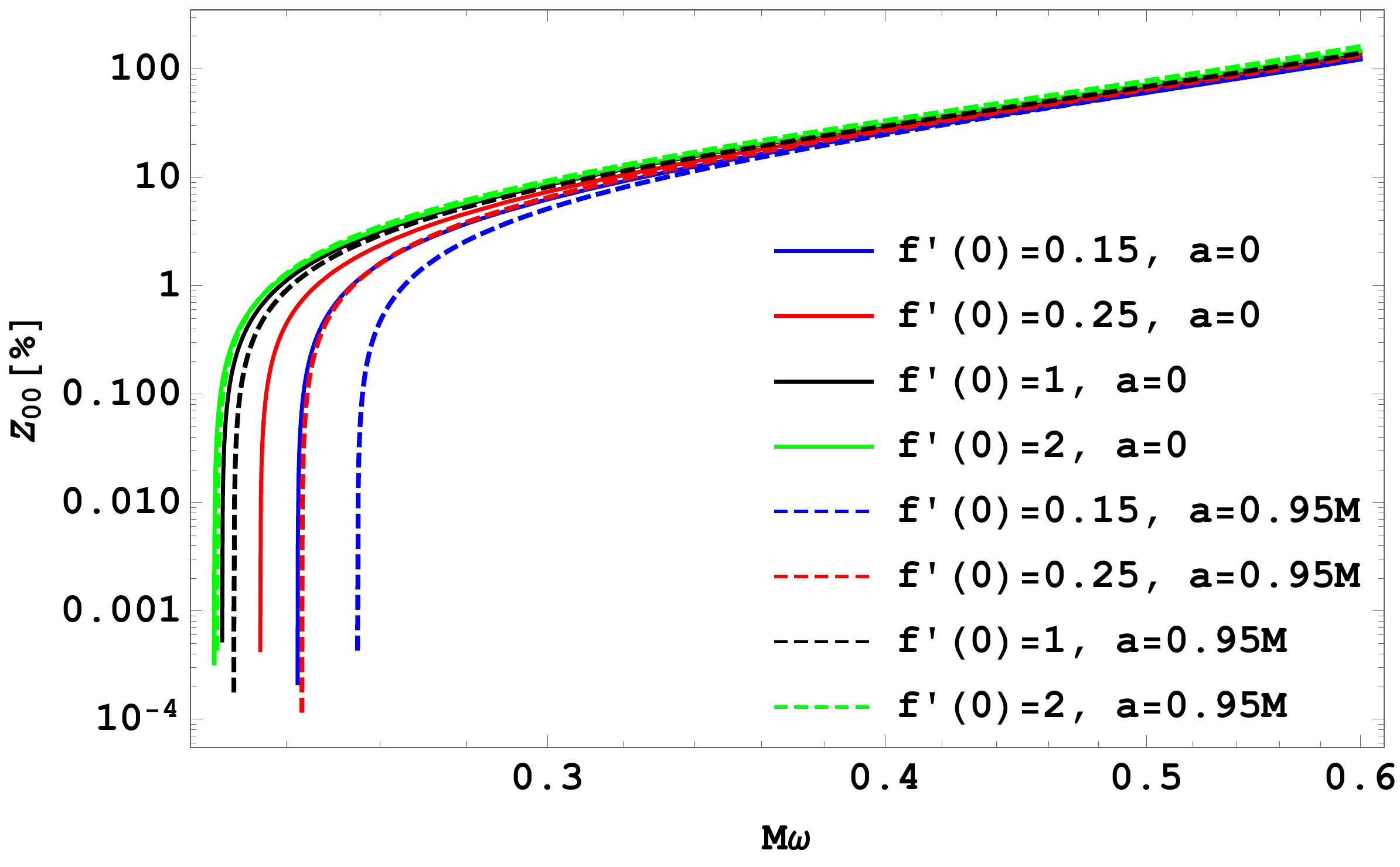}
	\includegraphics[scale=0.4]{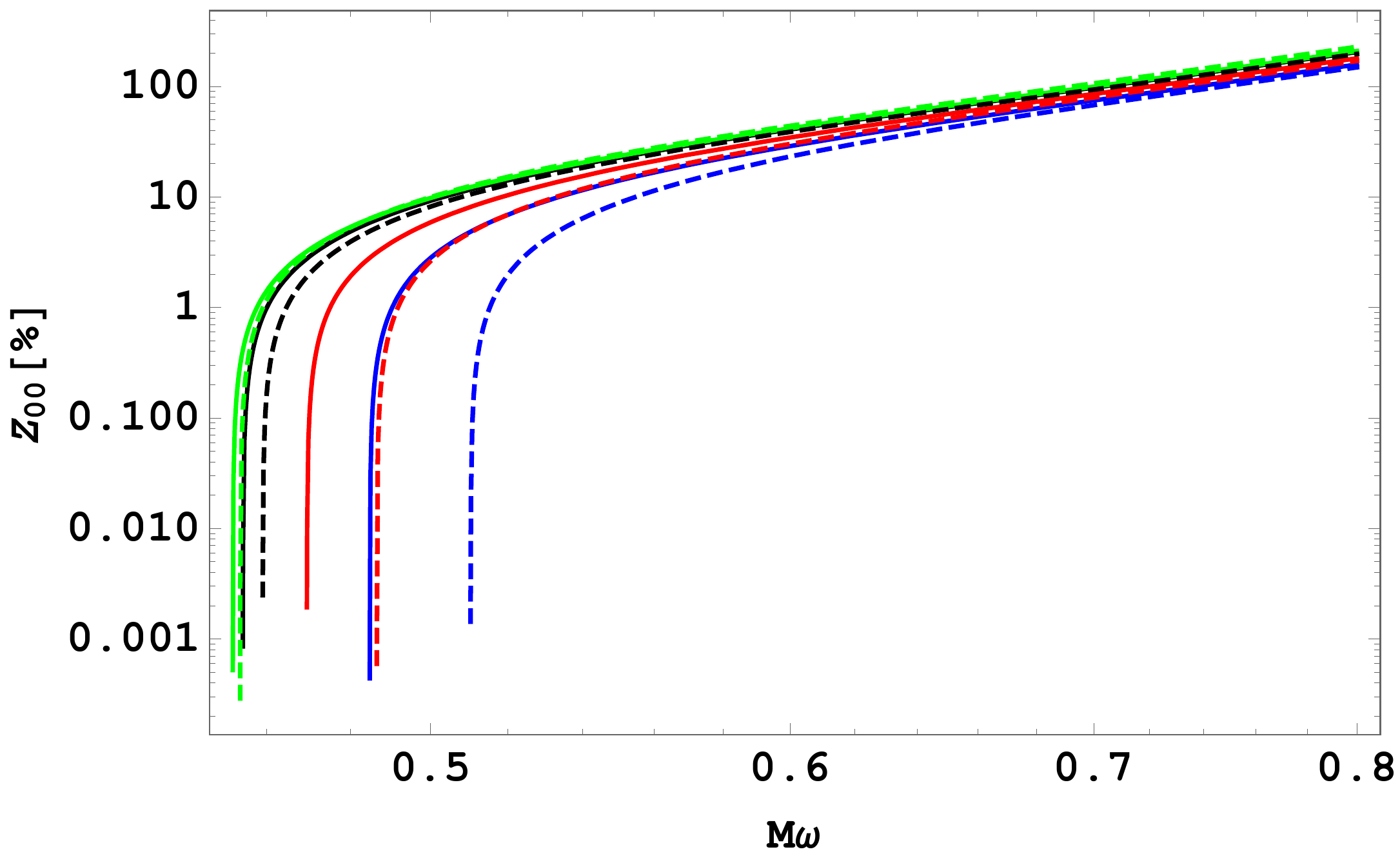}
	\includegraphics[scale=0.4]{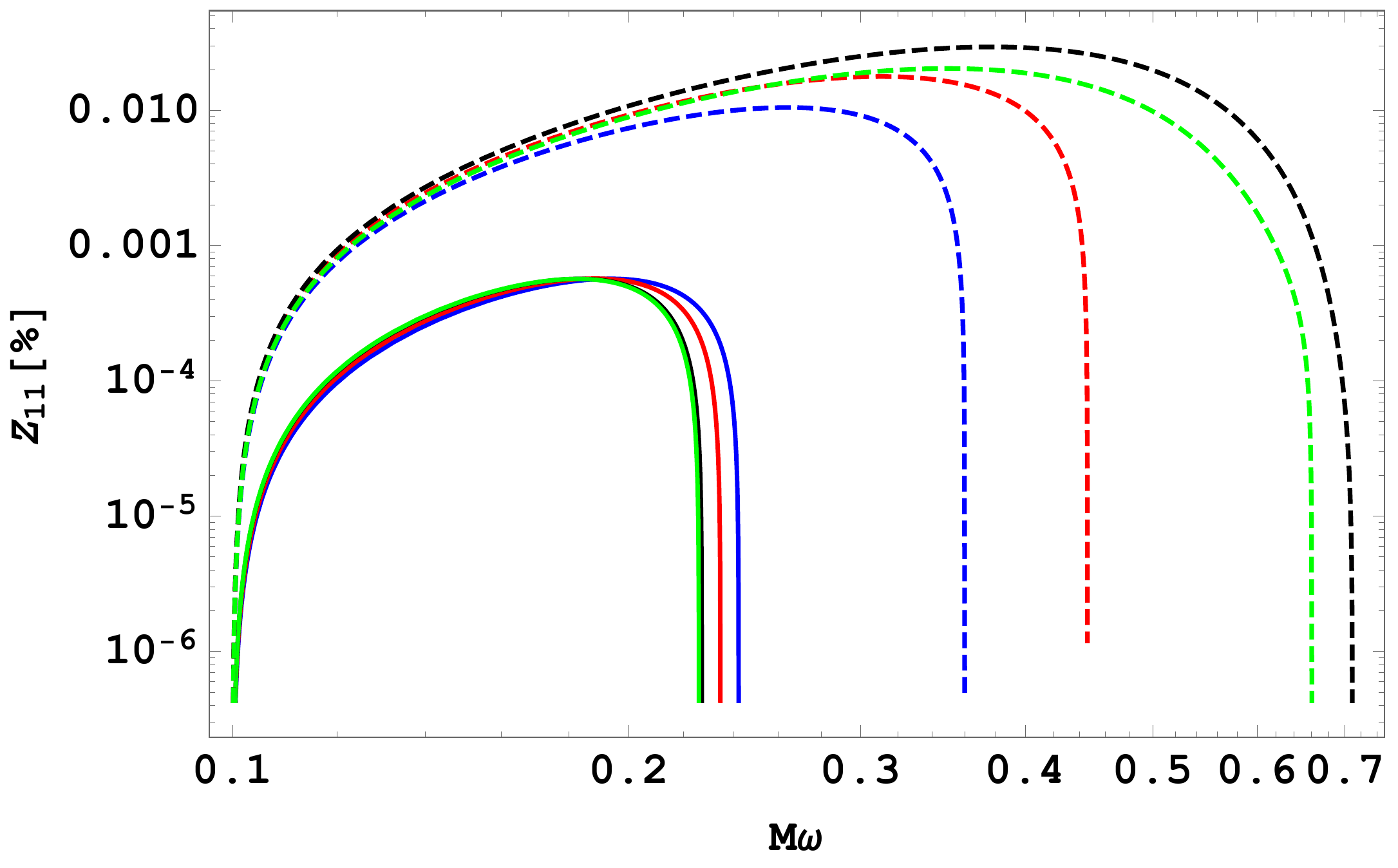}
	\includegraphics[scale=0.4]{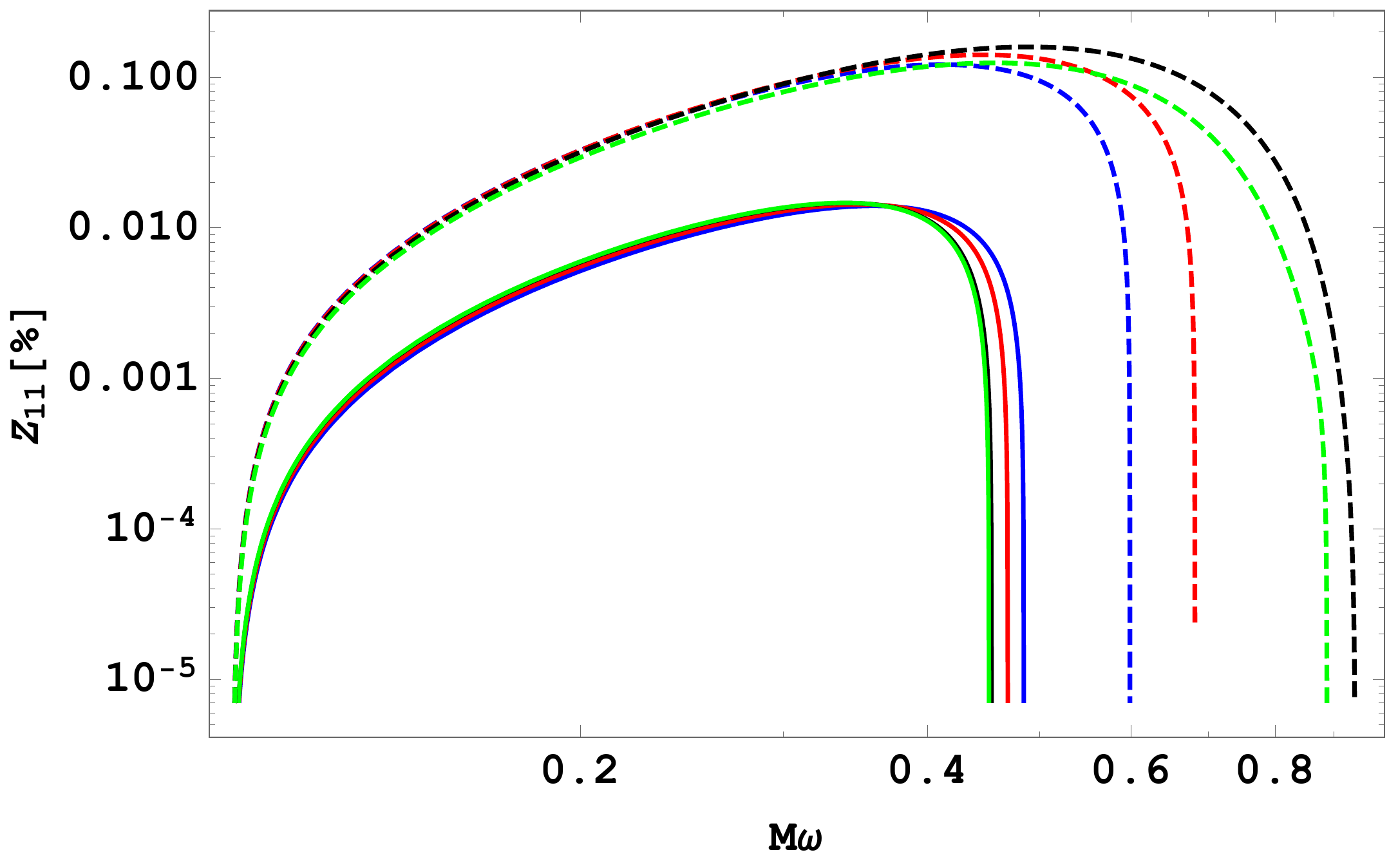}
    	\caption{ Percentage amplification factor $Z_{lm}$ (Eq. \eqref{amp}$\times 100$)
in terms of the frequency $\omega$ for a charged massive
scalar waves with mass  $\mu_{s}=0.1$ and modes: $l=0=m$ (top raw), $l=1=m$ (bottom raw) scattering off a $f(R)$-Kerr-Newman black hole with electric charge $q=0.1M$. The electric coupling $\varepsilon$
is $0.45$ (left panel) and $0.9$ (right panel).   }
\label{Amp1}
\end{figure}

 \begin{figure}[!ht]
 	\includegraphics[scale=0.4]{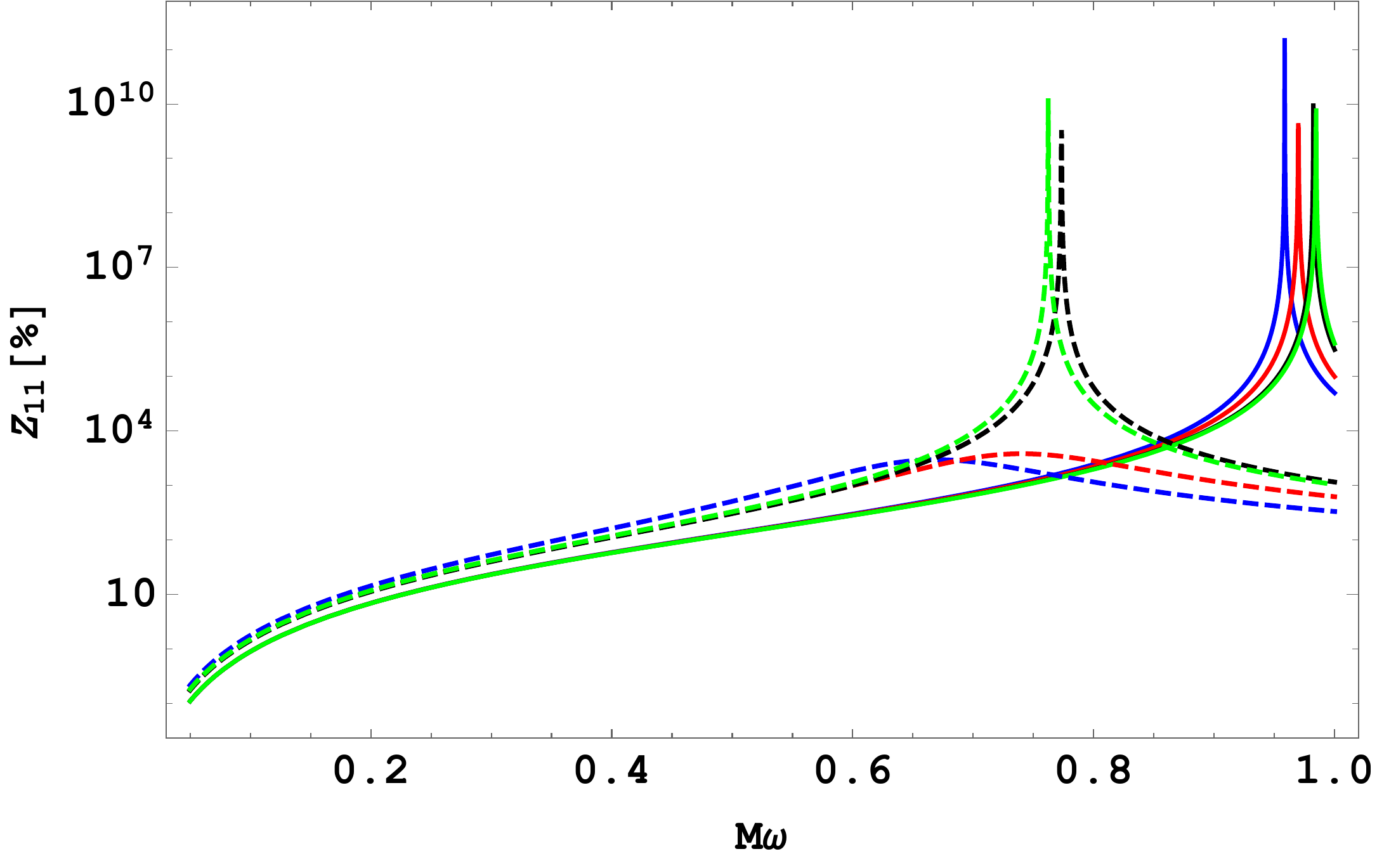}
	\includegraphics[scale=0.4]{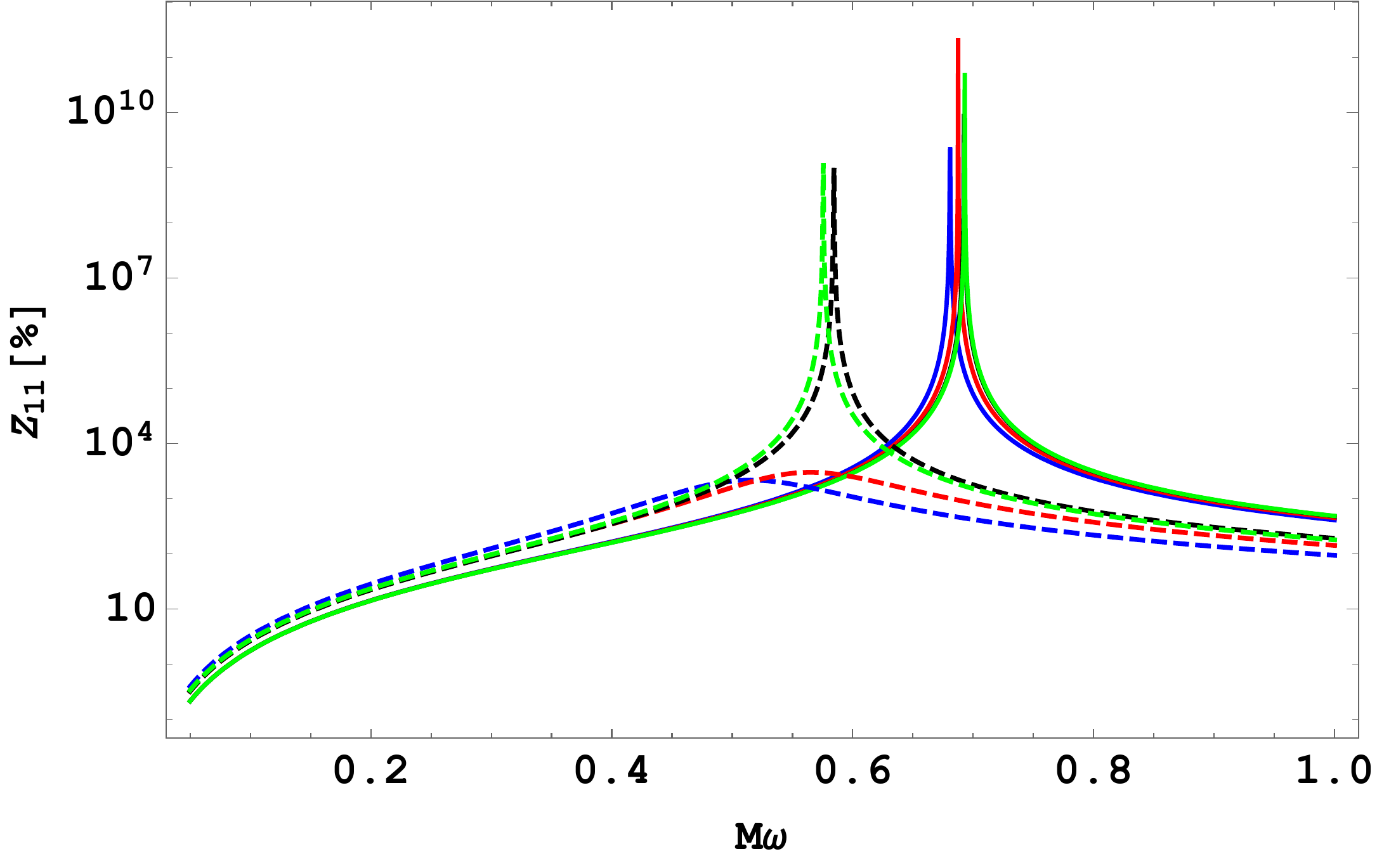}
    \caption{ Same as bottom row in the Fig. \ref{Amp1} but for large coupling regimes
    $\varepsilon=8$ (left panel) and  $\varepsilon=12$ (right panel).}
\label{Amp2}
\end{figure}

\subsubsection{Analysis with Viable $f(R)$ Models}
Now using the above expressions we are able to look for the effects  of $f(R)$ modified gravity
on the superradiant amplification. To do so, we study
two viable cosmological $f(R)$ models, as listed
in Table~\ref{f(r)} as our case studies.

In order to prevent the  ghost
and tachyonic instabilities  the following conditions are
required to be satisfied  for viable $f(R)$ theories \cite{Pogosian:2007sw}
\begin{eqnarray}\label{f}
f'(R_0)=\frac{d f}{dR}|_{R_0}>0,~~~~~f''(R_0)=
\frac{d^2 f}{dR^2}|_{R_0}\geq0\,,
\end{eqnarray}

The $f(R)$ models listed in Table~\ref{f(r)} satisfy  the conditions of
obtaining  the Kerr-Newman black holes in asymptotically flat
spacetime, i.e. $f(0)=0$ and $f'(0)\neq0$. Even though
model I explicitly satisfies both of the above stability conditions, but the satisfaction of these conditions for model II depends  on the  parameters $c_{1,2}$. In this model, to satisfy both stability conditions \eqref{f},
either of the following  three combinations of parameters should be satisfied
\begin{center}
$c_{1,2}$:  $\left( 0<c_1<1,~~c_2\geq0 \right) ;~~ \left(c_1<0,~~c_2\leq0 \right) ;~~  \left( c_1=0=c_2 \right) $~.
\end{center}

To have a view of the effects of $f(R)$ modification  on
the amplification factor, we have plotted the behaviour of  $Z_{lm}$ given in  Eq. \eqref{amp} in Figs. \ref{Amp1} and \ref{Amp2}. Generally speaking, we see the  amplification factors of ground state $(l=0=m)$
and the first excited state  $(l=1=m)$ are affected for values different from $f'(0)=1$. While rotation ($a\neq0$) has a negligible effect on the ground state mode with $f'(0)>1$, for the case of $0<f'(0)<1$ in this mode and also for all cases of $f'(0)$ for the mode $Z_{11}$  it has significant  effects on the power as well as the frequency range of superradiance. Focusing the Kerr-Newman black hole, Fig. \ref{Amp1} clearly shows that in some frequencies, $Z_{00}$ and $Z_{11}$  grow
as $f'(0)$ changes from $0.15$ to $1$. However, $Z_{11}$ for the case of $f'(0)>1$
is weaker than its standard counterpart.
Furthermore, the first excited state mode has a bigger superradiance parameter space
as $f'(0)$ moves from $0.15$ to $1$ while it becomes smaller for the ground state mode. Concerning the case of $f'(0)>1$,  the first excited state mode has a smaller superradiance parameter space relative to  $f'(0)=1$ while there is no significant impact on  the ground state mode.
As another interesting point, the behavior of these modes are
sensitive to the values of the electrical charge coupling $\varepsilon$. For both modes under consideration when approaching $\varepsilon=1$ the amplitude and the superradiance frequency range become bigger as $f'(0)$ moves from $0.15$ to $1$.
In  large coupling limit there appears some resonance peaks for the mode $Z_{11}$
in some  given frequencies $\omega_{res}$  as depicted in Fig. \ref{Amp2}.
The resonance frequencies $\omega_{res}$ as well as the amplitude of these
peaks grows as $f'(0)$ increases. However, we have found that the large coupling
regime does not support any solution for the case of the ground state mode $Z_{00}$.
This is not surprising  since, as already mentioned, one of the limitations of the AAM method is that it  requires weak coupling of the charged scalar field to Kerr-Newman black hole.

The shape of the resonance peaks is similar to Breit-Wigner (BW) form \cite{BW} since their
heights and widths are  respectively finite and very narrow with no infinities and zeros, as in
Dirac delta function. Such peaks have been interpreted  as the very long lived quasi-normal modes (corresponding to quasi-bound states) with $\omega_I\ll\omega_R $ so that $\omega_R\sim \omega_{res}$, see the discussions in \cite{Cardoso:2013opa}. Historically, the existence of such a long-lived quasi-normal modes  can be traced to the work of Detweiler \cite{Detweiler:1980gk}. However, for some more recent works on
these weakly damped quasi-normal modes see \cite{Berti:2009wx, Degollado:2013eqa, Richartz:2017qep}.
As a consequence, the BW-shaped resonances in Fig. \ref{Amp2} address the existence of
stable quasi-bound modes in asymptotically flat $f(R)$-Kerr-Newman black holes, specifically
in large coupling regime\footnote{It should be mentioned that the quasi-normal modes and scattering are two related phenomena in which  the resonant peaks are the poles of the scattering matrix  in the complex-frequency plane.}.

It should be noted that model I and the case $c_1=0$ in model II are similar to the case of GR with
$f'(0)=1$.   However, the cases $0<c_1<1,~~c_2\geq0$ and $c_1<0,~~c_2\leq0$
in model II can have  smaller or bigger values than  $f'(0)=1$.
As a result, from the two models listed in Table \ref{f(r)}, only in model II
with the aforementioned cases for $c_{1,2}$, the superradiance amplification as well as its
frequency ranges are distinguishable from those of GR. However, in the coming subsection we show that
the model I becomes distinguishable from GR if we place the black hole in an asymptotically de-Sitter spacetime.

 \begin{table}
	\begin{center}
		\begin{tabular}{|c|c|c|c|}
			\hline
			$f(R)$ models& Ref & $f'(R_0=\frac{12}{L^2})$ & $f''(R_0=\frac{12}{L^2})$   \\
			\hline
			\textbf{Model I:}~ $f=R+\alpha R^2,~~~\alpha>0$ & \cite{Starobinsky:1980te} &$ 1+\frac{24\alpha}{L^2}$ & $2\alpha$ \\ \hline
			\textbf{Model II:}~$f^{n=1}=R-\gamma^2\frac{c_1(R/\gamma^2)^n}{1+c_2(R/\gamma^2)^n}$&\cite{Hu:2007nk}&$\frac{ (1-c_1)\gamma ^4L^4+24 c_2 \gamma ^2 L^2+144 c_2^2}{\left(12 c_2+\gamma ^2 L^2\right)^2}$&$\frac{2  c_1 c_2 \gamma ^4 L^6}{\left(12 c_2+\gamma ^2 L^2\right)^3}$
			\\  \hline
			\end{tabular}
		\caption{Two viable cosmological $f(R)$ models in asymptotically de-Sitter spacetime. }
		\label{f(rr)}
	\end{center}
\end{table}

\subsection{Asymptotically de-Sitter spacetime: $R_0>0$}

Given the fact that the de-Sitter spacetime is bounded (observer can not see beyond the cosmological horizon $r_H$) so now the near and far regions
approximations are  altered to $r_H-r_h\ll\omega^{-1}$ and $r_H-r_h\gg M$ respectively.
To maintain the validity of these approximations, we have to assume that the
cosmological horizon  is also much larger than  the event horizon, $r_H\gg r_h$. This
means that  for the spacetime near the vicinity of the
charged Kerr black the  large limit of
the near-region  solution in Eq. \eqref{solnear} is applicable here. So,
in the following, we  focus on the solution in the far regions.
Inspired by the far region approximation $r_H-r_h\gg M$, the black hole's
mass, electric charge and angular momentum parameters do not play important roles and can be
ignored for a distant observer  near to the cosmological horizon ($M\sim0,~Q\sim0,~a\sim0$).
Namely, $\Delta_r\approx r^2(1-\frac{r^2}{L^2})$
where $L$ is the radius of the de-sitter spacetime  defined as $L\equiv \sqrt{\frac{12}{R_0}}$.

\begin{figure}[!ht]
	\includegraphics[scale=0.32]{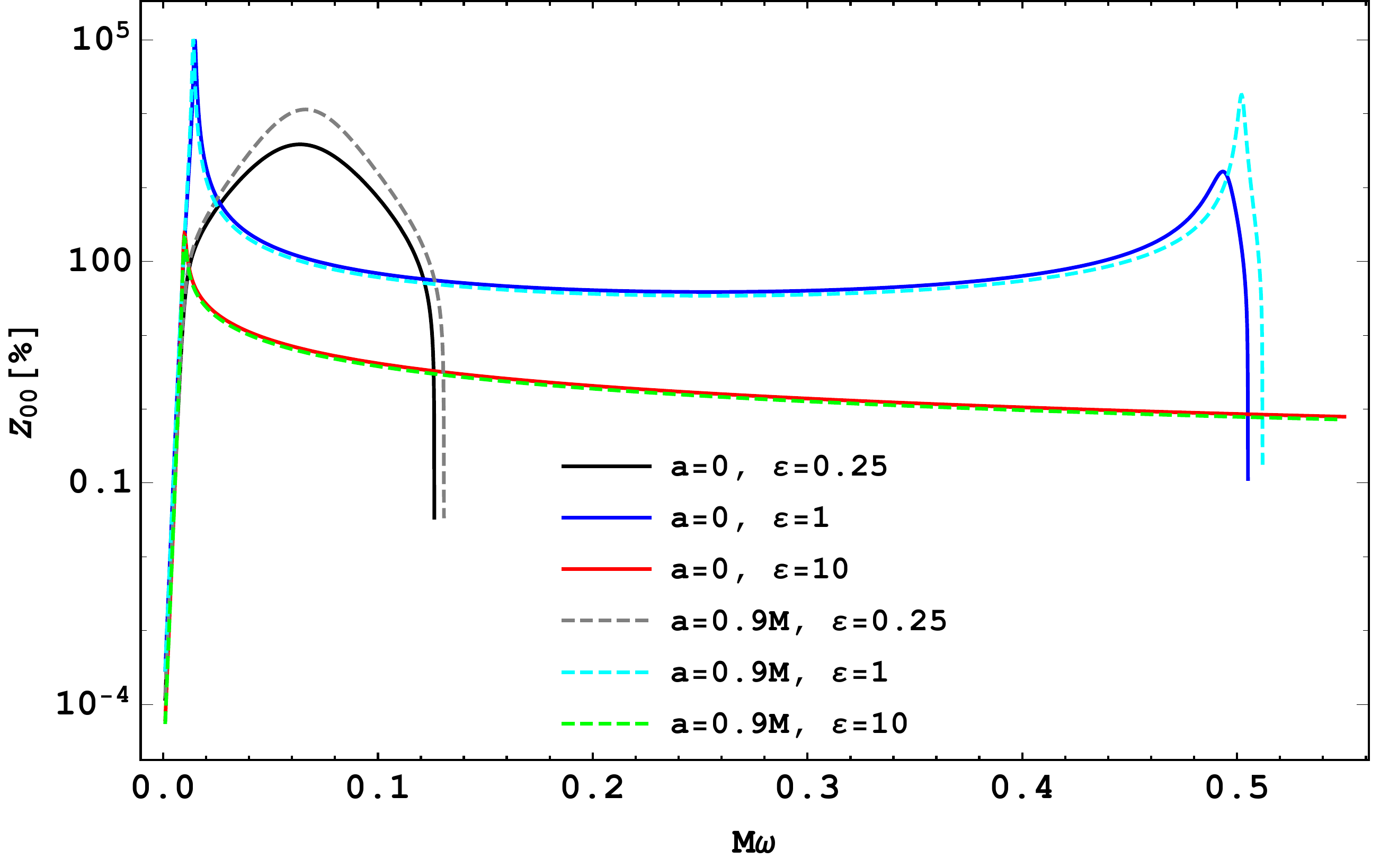}
	\includegraphics[scale=0.32]{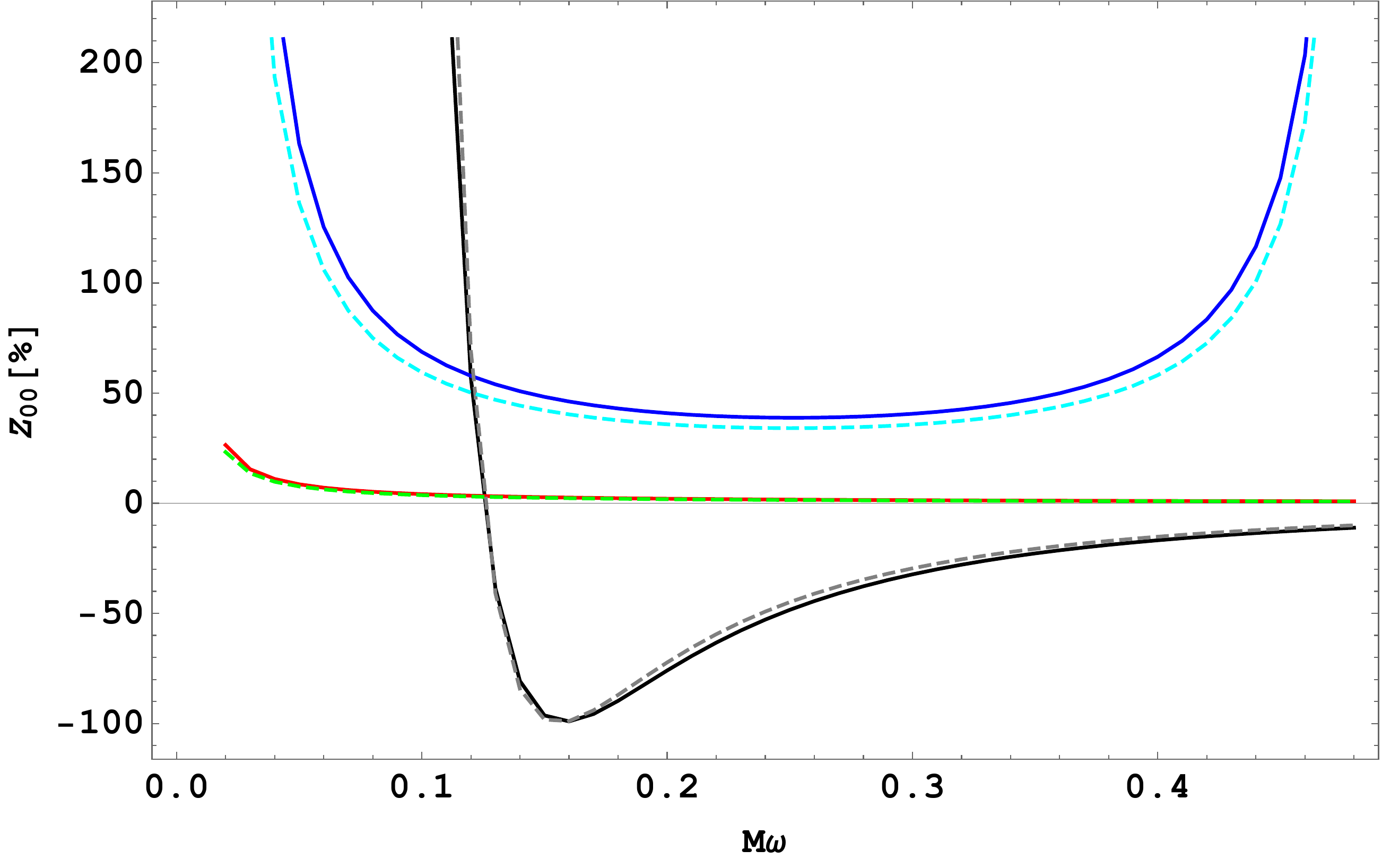}
    \includegraphics[scale=0.32]{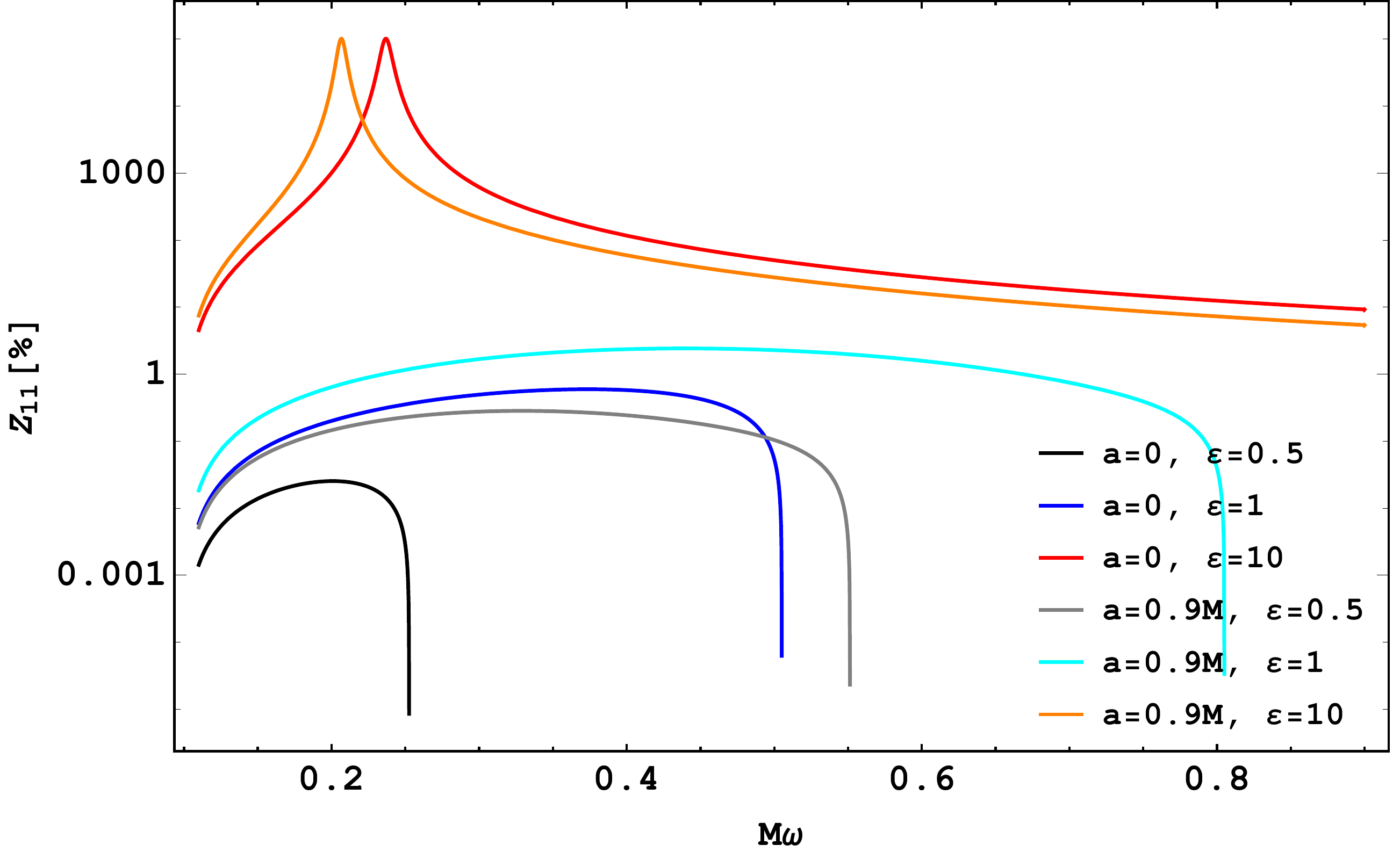}
    \includegraphics[scale=0.32]{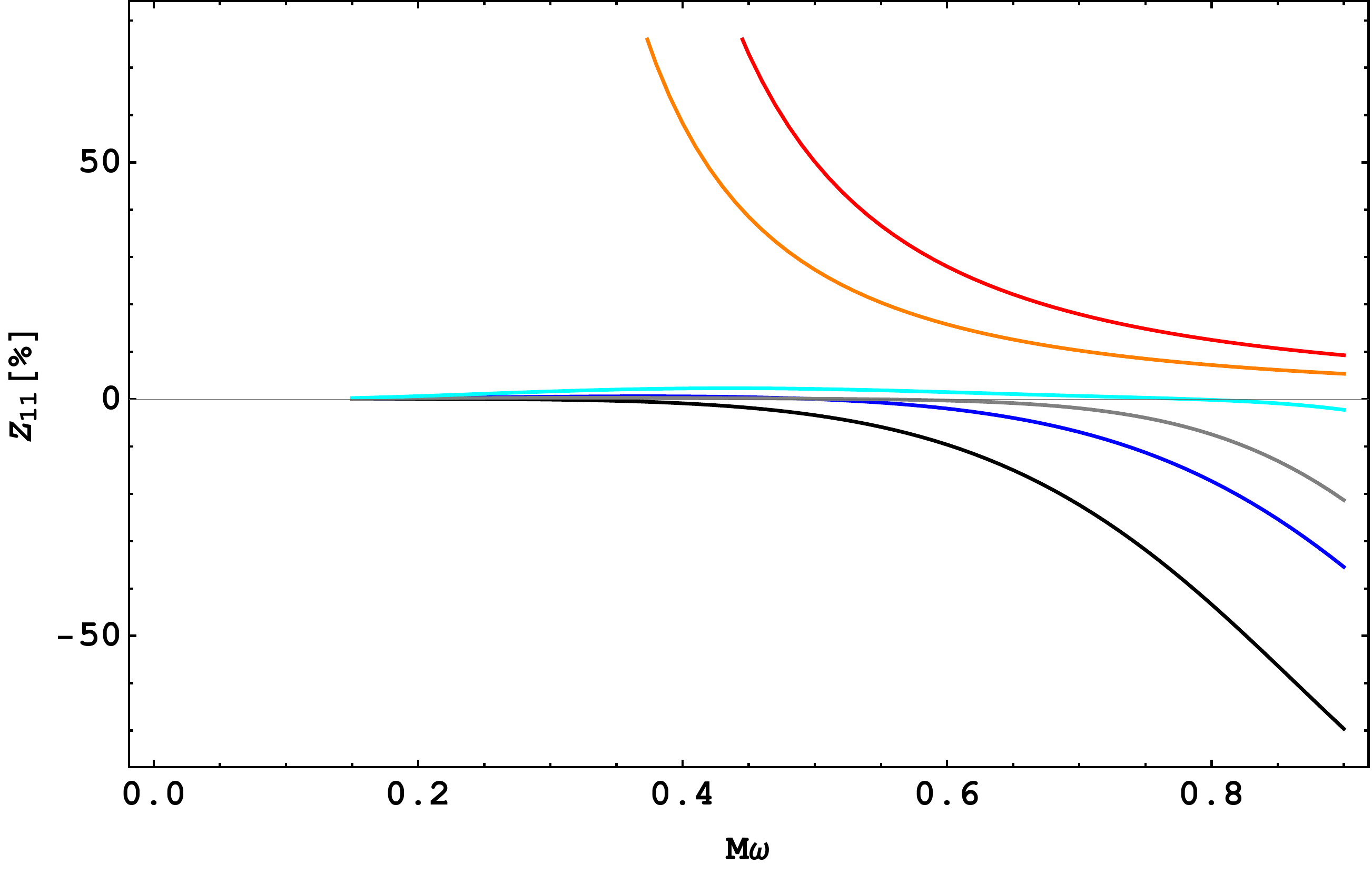}
	\caption{ Percentage amplification factor $Z_{lm}$ (Eq. \eqref{far9-ds}$\times 100$) vs.  the frequency $\omega$ for $\mu_s=0.1$ and modes: $l=0=m$ (top row) and $l=1=m$ (bottom row) scattering
 off a GR-Kerr-Newman-dS black hole (left panels) and $f(R)$-Kerr-Newman-dS black hole in model I (right panels) with the electrical charge $q=0.1M$. For the de-Sitter length we set value $L=500$.
\vspace{1cm}
}
		\label{Amp3}
        \end{figure}

        \begin{figure}[!ht]
	\includegraphics[scale=0.33]{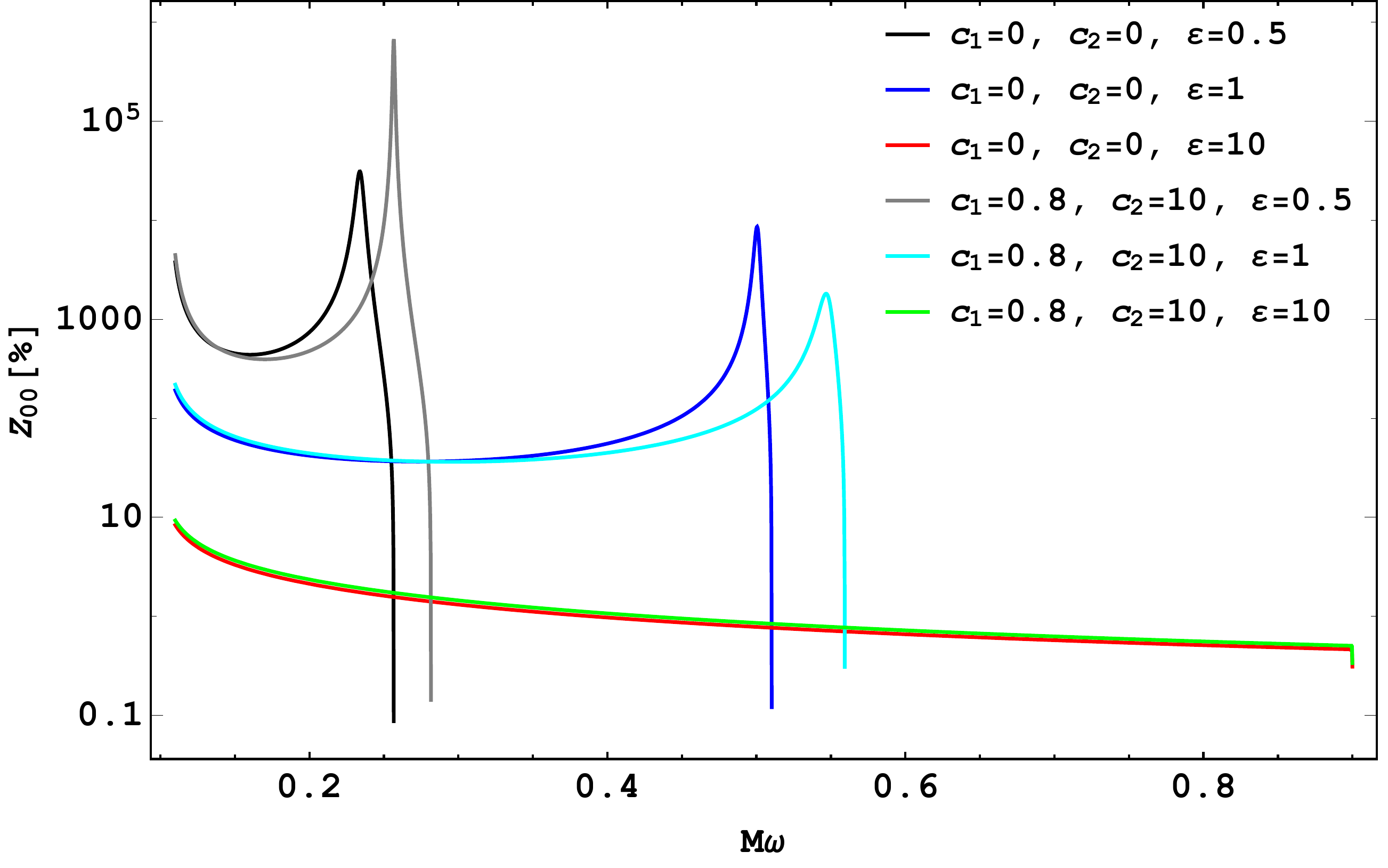}
	\includegraphics[scale=0.33]{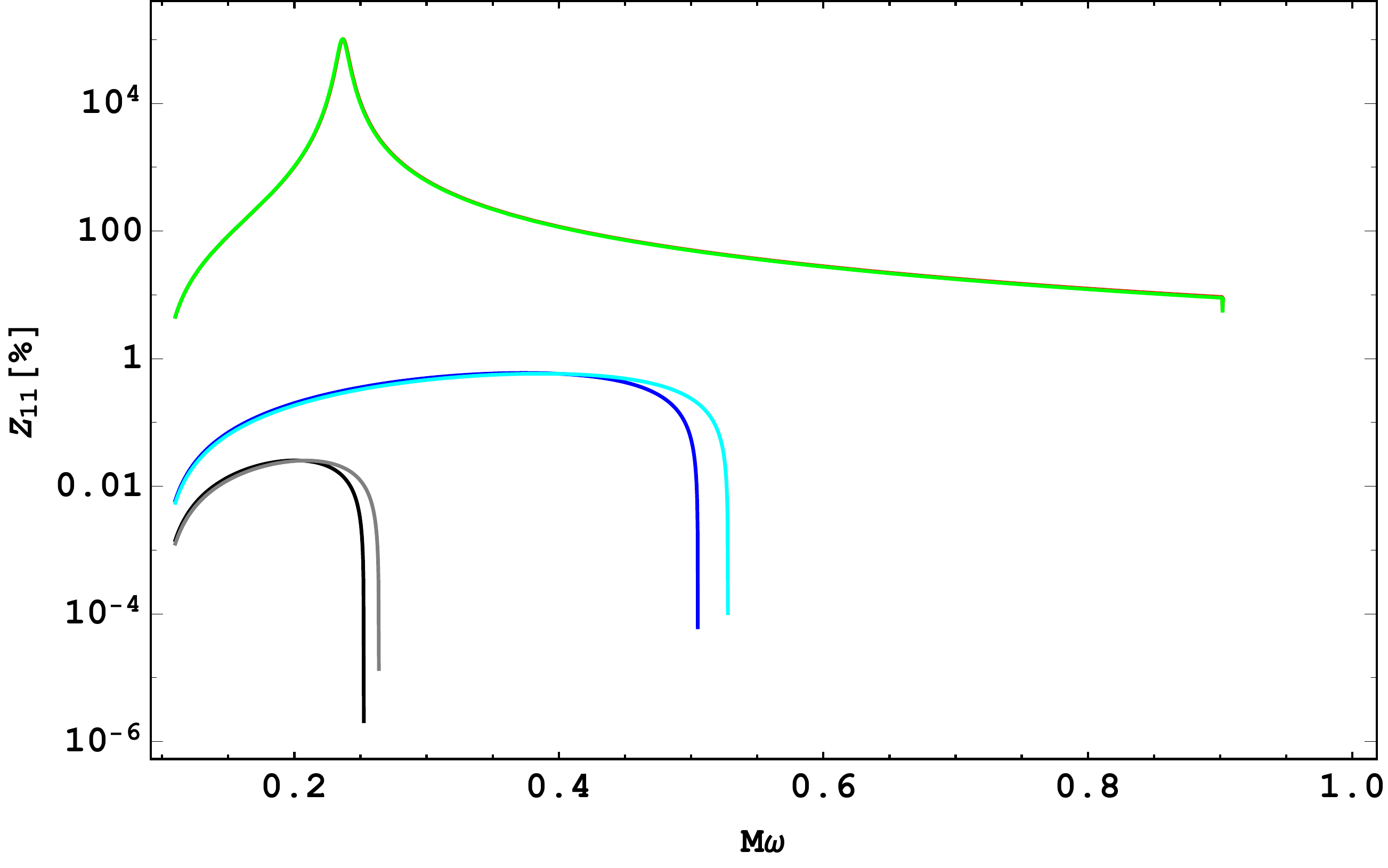}
	\caption{ Same as Fig.~\ref{Amp3} but for $f(R)$ model II. The general behaviour of plots do not change for any value of $c_2>0$  and $\gamma>0$ or $\gamma<0$ (here we set $\gamma=20$). The curves with $c_1=0=c_2$ have the same predictions  as in GR. 
	}
	\label{Amp4}
\end{figure}
Now, the radial wave equation  \eqref{KGR} represents the  propagation
of a charged massive scalar field (with frequency $\omega$ and angular momentum $l$) in a pure dS
spacetime,
\begin{eqnarray}\label{far-ds}
\Delta_r\frac{d^2\mathcal{R}}{dr^2}+\frac{d\Delta_r}{dr}\frac{d\mathcal{R}}{dr}+\bigg[\frac{\omega^2r^4}{\Delta_r}-\bigg(l(l+1)
+\mu_s^2r^2\bigg)\bigg]\mathcal{R}=0~.
\end{eqnarray}
In terms  of the new coordinates $y\equiv 1-\frac{r^2}{L^2}$ and the new field variable
\begin{eqnarray}\label{ch-ds}
\mathcal{R} \equiv y^{\frac{iL\omega}{2}}(1-y)^{l/2}G(y)~,
\end{eqnarray}
the differential equation \eqref{far-ds} is cast into
\begin{eqnarray}\label{far2-ds}
y(1-y)\frac{d^2G}{dy^2}+\big((1+i L \omega)-(l+\frac{5}{2}+i L\omega)y\big)\frac{dG}{dy}
-\frac{1}{4}\big((l+iL\omega)(l+3+iL\omega)+L^2\mu_s^2\big)G=0.
\end{eqnarray}
Defining the following parameters
\begin{eqnarray}\label{far3-ds}
\beta_1=1+iL\omega,~~~\beta_{2,3}=\frac{3\mp\sqrt{9-4L^2\mu^2}+2(l+iL\omega)}{4}
~,
\end{eqnarray}
the radial equation takes the  well known Gaussian hypergeometric
differential equation,
\begin{eqnarray}\label{far4-ds}
y(y-1)\frac{d^2G}{dy^2}+\bigg((\beta_2+\beta_3+1)y-\beta_1\bigg)\frac{dG}{dy}+\beta_2\beta_3G=0~.
\end{eqnarray}
For non-integer  $\beta_1$,  the general solution of the differential equation  can be written
as \cite{book-dif}
\begin{eqnarray}\label{far5-ds}
\mathcal{R}=(1-y)^{l/2} \Big[
D_1y^{\frac{iL\omega}{2}}\, _{2}F_1\big(\beta_2,\beta_3,\beta_1;y\big)+
D_2y^{\frac{-iL\omega}{2}}\, _{2}F_1\big(\beta_2-\beta_1+1,\beta_3-\beta_1+1,2-\beta_1;y\big) \Big]\, ,
\end{eqnarray} with constant coefficients $D_{1,2}$.

Considering the small $r$ limit, i.e. $y\rightarrow1~$,  the solution is
\begin{align}\label{far6-ds}
\mathcal{R}=\big(B_1D_1+C_1D_2\big)r^l+\big(B_2D_1+C_2D_2\big)r^{-l-1}~,
\end{align}
with
\begin{align}\label{far7-ds}
&B_1=-\frac{\pi \Gamma(\beta_1)}{L^l\sin\bigg((\beta_2+\beta_3-\beta_1)\pi\bigg)\Gamma(1+\beta_2+\beta_3-\beta_1)\Gamma(\beta_1-\beta_2)\Gamma(\beta_1-\beta_3)} ,\nonumber\\
&C_1=-\frac{\pi \Gamma(2-\beta_1)}{L^l\sin\bigg((\beta_2+\beta_3-\beta_1)\pi\bigg)\Gamma(1+\beta_2+\beta_3-\beta_1)\Gamma(1-\beta_2)\Gamma(1-\beta_3)} ,\nonumber\\
&B_2=\frac{\pi \Gamma(\beta_1)L^{l+1}}{\sin\bigg((\beta_2+\beta_3-\beta_1)\pi\bigg)\Gamma(\beta_3)\Gamma(\beta_2)\Gamma(1+\beta_1-\beta_2-\beta_3)}, \nonumber\\
&C_2=\frac{\pi \Gamma(2-\beta_1)L^{l+1}}{\sin\bigg((\beta_2+\beta_3-\beta_1)\pi\bigg)\Gamma(1+\beta_2-\beta_1)\Gamma(1+\beta_3-\beta_1)\Gamma(1+\beta_1-\beta_2-\beta_3)} .
\end{align}
Now by matching the solution \eqref{far6-ds} with solution \eqref{solnear}, we can solve for  the coefficients $D_{1,2}$ as
\begin{eqnarray}\label{far8-ds}
D_1=\dfrac{C_2\alpha_1-C_1\alpha_2}{B_1C_2-B_2C_1},~~~~~~~
D_2=\dfrac{B_2\alpha_1-B_1\alpha_2}{B_2C_1-B_1C_2}~,
\end{eqnarray}
where $\alpha_{1,2}$ are obtained from  \eqref{solnear} as follows
\begin{eqnarray}
\alpha_1=C(r_h-r_c)^{-l}
\frac{\Gamma(2l+1)\Gamma(1-2i\eta)}{\Gamma(l+1-2i\eta)\Gamma(l+1)}, \nonumber\\
\alpha_2=C(r_h-r_c)^{l+1}
\frac{\Gamma(-2l-1)\Gamma(1-2i\eta)}{\Gamma(-l)\Gamma(-l-2i\eta)}.
\end{eqnarray}

As the last step, we have to expand the equation \eqref{far5-ds} around the cosmological horizon $r_H$,  i.e. $y\longrightarrow0$ and subsequently compare it with the solution \eqref{ds-so} on $r_H$. As a result, the amplification factor is given by
\begin{eqnarray}\label{far9-ds}
Z_{lm}=\frac{|D_1|^2}{|D_2|^2}-1~.
\end{eqnarray}

The amplification factors $Z_{00}$ and $Z_{11}$ for the $f(R)$ model I are displayed in Fig.~\ref{Amp3}.
Note that the $f(R)$ model I (the Starobinsky model) is well consistent with cosmological observations for early universe cosmology and inflation with the parameter
 $\alpha\sim 10^{10}$ \cite{Salehian:2018yoq}. {Interestingly, as displayed in Fig.~\ref{Amp3},
 we find that for the mentioned value of $\alpha$ there is a weak chance of  superradiance since generally both $Z_{00}$ and $Z_{11}$ are negative or small positive values (compared with GR) in the limit of our interest where $M\omega\ll 1$.  It can even be seen that the expected resonance peaks in GR for the case of $\varepsilon>1$ are disappeared here. As a result, we conclude that
for the Kerr-Newman black hole in a de-Sitter background the Starobinsky model with the required value of parameter $\alpha$ does not  support the superradiance phenomena.}

Our analysis show that for the   $f(R)$ model II  superradiance can be either enhanced or reduced.
Note that in the case of $c_1=0=c_2$ the results of this model coincides with those of GR,  independent of the value of $\gamma$.
Clearly one can see from Table \ref{f(rr)} that for values $0<c_1\leq1,~c_2>0$ with any arbitrary
value for $\gamma$ the required  conditions $f'(R_0)>0$ and $f''(R_0)>0$  are always satisfied. As revealed in Fig.~\ref{Amp4},
the behavior of SAFs for $Z_{00}$ and $Z_{11}$ are different from what is shown for the model I in Fig.~\ref{Amp3}. {For the ground state mode $Z_{00}$ we see that the superradiance is weakened as $\varepsilon$ is close to unity but the frequency parameter space becomes wider compared to the case of GR.
Also,  similar to GR, one can see some resonance peaks at the at the end of the superradiance parameter space.  However, for the mode $Z_{11}$, the resonance peak appears only in the large coupling regime. The  frequency of this  resonance peak is indistinguishable from the case of GR while its amplitude is equal to the case of GR. For the rest the amplitude and the parameter space of superradiance
are partly stronger and wider than in the case of GR respectively.} An interesting observation from the  above figures is the existence of solutions for the large coupling regions $\varepsilon>1$ but $M\omega<1$. However, these solutions may not be trusted as they deviate from the condition of the applicability  of
the AAM method.

\section{Summary and Conclusion}\label{Sum}

The details of the black hole superradiance amplification depend both on the geometry of black hole and the wave dynamics in the modified theories of gravity. Therefore, it is an interesting question to study the superradiance phenomenon in modified theories of gravity. In this work, we have studied this question for a charged massive scalar  wave scattering off small and slow rotating $f(R)$-Kerr-Newman black holes in asymptotically flat and de-Sitter spacetimes respectively. While our analysis were general, but as case studies we have presented our results for two $f(R)$ models, the Starobinsky model
\cite{Starobinsky:1980te} and the Hu-Sawicki model \cite{Hu:2007nk}. The main feature that distinguishes this black hole solution from its standard counterpart is that here the contribution of the black hole's charge to the metric carries an additional effect given by the factor $\frac{1}{\sqrt{f'(R_0}}$. We have argued  that
this extra effect is not degenerate  with the black hole's electric charge so it leaves distinguishable imprints in $f(R)$ models of gravity. Alternatively,  the corrections arising from $f(R)$ may be viewed as the change in the effective gravitational constant $G_{eff}=\frac{G_N}{f'(R_0)}$. 

We have found that the induced curvature correction  affects the underlying phenomenon so black hole superradiance scattering may provide a platform to distinguish GR from $f(R)$ theories.  Below we summarize our results  for the cases of
asymptotically flat and dS spacetimes separately.

 \begin{itemize}
  \item \textbf{Asymptotically flat spacetime:}

In the case of  asymptotically flat spacetime  we have found that only in $f(R)$ model II the superradiance amplification is distinguishable from those of GR with $f'(0)=1$. { In the weak coupling limit
$\varepsilon\leq1$ our analysis explicitly show that for the ground state as well as  the first excited modes we have a bigger superradiance parameter space with stronger amplitude as $f'(0)$ moves from $0.15$ towards larger values. It  should be noted that for  $f'(0)>1$ in the first excited modes the frequency range become smaller while  the amplitude become larger than GR.
In the large coupling regime $\varepsilon>1$, we have observed some Breit-Wigner shaped resonances describing the quasi-bound states with their peak frequencies increasing as $f'(0)$ changes from $0.15$ towards larger values. Generally, though the SAFs in the Starobinsky $f(R)$ model I are not distinguishable from those of GR, but depending on model parameters, SAFs in model II could be weaker or stronger than in GR.}

  \item \textbf{Asymptotically dS spacetime:}\\
In the case of asymptotically dS spacetime the predictions of superradiance scattering in both $f(R)$ models are different from the predictions of the asymptotically flat spacetime. We have shown that in the Starobinsky model with the free parameter $\alpha$ in the range to be consistent with the inflationary predictions either the model does not support superradiance or has a negligible chance compared with GR. However, in the Hu-Sawicki model the amplitudes
as well as the frequency ranges are different from those of GR. In this model, we also have seen some resonance peaks in SAFs corresponding to the quasi-bound states.  These peaks appear at different frequencies than in GR which may have interesting astrophysical implications.

\end{itemize}

At the end, it is necessary to point out three issues. First, despite the existence of some solutions for the large coupling regions $\varepsilon>1$ and $M\omega<1$, these solutions are not trusted  as they deviate from the limit of the applicability of the AAM method.
Second, the difference in the range of
superradiance frequency between $f(R)$ gravity and GR can have phenomenological importance. Indeed,
there are  some ranges of frequency in which GR does not support superradiance while superradiance
occurs in $f(R)$ gravity. This can be reversed as well, i.e.  there are frequency ranges in which
$f(R)$ does not support superradiance while it occurs in GR. As a result the  shift in superradiance
regime relative to GR may be an important observational tool to distinguish between GR and $f(R)$ theories.
In particular, as shown in  \cite{Richartz:2013hza}, stars in GR are capable of
superradiance amplification. Naturally, one expects that superradiance amplification to occur in astrophysical phenomena in $f(R)$ theories which may be distinguishable from the GR cases.
Third, we comment that since $f(R)$ theories are equivalent to the  generalized Brans-Dicke gravity the results obtained here can
be viewed as spacial cases of  the general scalar-tensor theories.

\vspace{0.5cm}
{\bf Acknowledgments:}
We would like to thank Carlos Herdeiro  for insightful comments and discussions and Mohammad Hossein Namjoo for
discussions and collaboration at the early stage of this work.

{}

\end{document}